\documentclass[aip,jcp,amsmath,amssymb,reprint,nolongbibliography]{revtex4-2}

\usepackage[a-1b]{pdfx} 

\usepackage{graphicx}
\usepackage{dcolumn}
\usepackage{bm}

\usepackage[utf8]{inputenc}
\usepackage[T1]{fontenc}
\usepackage{mathptmx}
\usepackage{listings}
\usepackage{rotating} 

\usepackage{color}
\usepackage{dcolumn} 
\usepackage{bm} 
\usepackage{graphicx}
\usepackage{multirow} 
\usepackage{pifont} 
\usepackage{epsfig}
\usepackage{amsmath} 
\usepackage{amssymb}
\usepackage{subfigure}
\usepackage{float}
\usepackage{booktabs}
\usepackage{tabularx}
\usepackage[sort&compress]{natbib}
\usepackage{gensymb}
\usepackage{rotating}
\usepackage{threeparttable}
\usepackage{comment}
\usepackage{physics}
\usepackage[normalem]{ulem}
\usepackage{hyperref} 
\hypersetup{
  colorlinks,
  linkcolor=blue,
  citecolor=blue,
  urlcolor=blue
} 

\usepackage{footmisc} 
\begin{document}

\author{Stephen H. Yuwono}
\affiliation{
       Department of Chemistry and Biochemistry,
       Florida State University,
       Tallahassee, FL 32306-4390, USA}

\author{Run R. Li}
\affiliation{
       Department of Chemistry and Biochemistry,
       Florida State University,
       Tallahassee, FL 32306-4390, USA}

\author{Tianyuan Zhang}
\affiliation{Department of Chemistry, University of Washington, Seattle, WA 98195, USA}

\author{{\color{black} Kshitijkumar A. Surjuse}}
\affiliation{Department of Chemistry, Virginia Tech, Blacksburg, VA 24061, USA}

\author{Edward F. Valeev}
\affiliation{Department of Chemistry, Virginia Tech, Blacksburg, VA 24061, USA}
       
\author{Xiaosong Li}
\email{xsli@uw.edu}
\affiliation{Department of Chemistry, University of Washington, Seattle, WA 98195, USA}

\author{A. Eugene DePrince III}
\email{adeprince@fsu.edu}
\affiliation{
       Department of Chemistry and Biochemistry,
       Florida State University,
       Tallahassee, FL 32306-4390, USA}

\title{Relativistic coupled cluster with completely renormalized and perturbative triples corrections}



\begin{abstract}

We have implemented noniterative triples corrections to the energy from coupled-cluster with single and double excitations (CCSD) within the 1-electron exact two-component (1eX2C) relativistic framework. The effectiveness of both the CCSD(T) and the completely renormalized (CR) CC(2,3) approaches are demonstrated by performing all-electron computations of the potential energy curves and spectroscopic constants of copper, silver, and gold dimers in their ground electronic states. Spin-orbit coupling effects captured via the 1eX2C framework are shown to be crucial for recovering the correct shape of the potential energy curves, and the correlation effects due to triples in these systems changes the dissociation energies by about 0.1--0.2 eV or about 4--7\%. We also demonstrate that relativistic effects and basis set size and contraction scheme are significantly more important in Au$_2$ than in Ag$_2$ or Cu$_2$.
\end{abstract}

\maketitle

\section{Introduction}

\label{SEC:INTRODUCTION}

Single-reference coupled-cluster (CC)\cite{Coester58_421,Cizek66_4256,Cizek69_35,Paldus71_359} theory has established itself as one of most successful approaches for high-accuracy \emph{ab initio} electronic structure calculations. The exponential {\em ansatz} defining the ground-state CC wave function, combined with the connected\cite{Brueckner55_36,Goldstone57_267,Hubbard57_539,Hugenholtz57_481} and linked\cite{Hubbard57_539,Hugenholtz57_481} cluster theorems for the energy and wave function, respectively, lead to desirable properties such as size-extensivity and size-consistency of the ground-state CC energetics. Systematic truncation of the cluster operator leads to the well-known hierarchy of methodologies starting from CC with singles and doubles (CCSD)\cite{Bartlett82_1910,Zerner82_4088} to CC with singles, doubles, and triples (CCSDT),\cite{Bartlett87_7041,Schaefer88_382} CC with singles, doubles, triples, and quadruples (CCSDTQ),\cite{Adamowicz91_6645,Bartlett91_387,Bartlett92_4282,Adamowicz94_5792} and so on. It is well known that the CCSD, CCSDT, CCSDTQ, {\em etc.}~series provides a rapid convergence toward full configuration interaction (CI); the basic CCSD approach usually provides a qualitatively correct description of the system, and, with the inclusion of triples or quadruples, one can obtain a converged result relative to full CI.\cite{Musial07_291}

Although conceptually straightforward, the inclusion of higher-order cluster operators comes with a steep increase in the computational effort. Indeed, the computational effort of a CCSD calculation, using a properly factorized implementation, scales as {\color{black}$\mathcal{O}(n_o^2 n_u^4)$}, where $n_o$ and $n_u$ denote the numbers spin orbitals that are occupied and unoccupied in the reference configuration, respectively. This scaling increases to {\color{black}$\mathcal{O}(n_o^3 n_u^5)$} for CCSDT and {\color{black}$\mathcal{O}(n_o^4 n_u^6)$} for CCSDTQ, rendering such high-level CC calculations impractical for large molecules or basis sets. Thus, one of the primary goals in the development of new CC methodologies is the incorporation of correlation effects due to higher--than--doubly--excited cluster operators without incurring a significant increase in the computational cost. Efforts in this area have resulted in a variety of approaches, such as the inclusion of selected parts of high-order cluster components in an iterative manner,\cite{Urban87_126,Jorgensen95_7429,Helgaker97_1808,Bartlett99_6103,Piecuch10_2987} non-iterative corrections to low-order CC
{\color{black} energetics,\cite{HeadGordon89_479,Stanton97_130,Stanton98_601,Bartlett98_5243,Bartlett98_5255,Bartlett08_044110,Bartlett08_044111,Piecuch04_12197,Wloch05_224105,Gour06_2149,Kinal06_467,Piecuch07_11359,Wloch08_2128,Piecuch12_180,Piecuch12_144104,Piecuch12_4968}}
and tensor decomposition techniques,\cite{Lesiuk20_453,Lesiuk21_7632} to name a few examples. Among these options, non-iterative corrections have been a popular choice due to the relative simplicity of their implementation, especially if one is not interested in correcting the CC wave function itself.

Focusing on triples correlation effects, the CCSD(T)\cite{HeadGordon89_479} approximation to CCSDT has been hailed as the ``gold standard'' due to its ability to capture a significant part of the desired correlations, especially near equilibrium geometries. However, CCSD(T) was derived using many-body perturbation theory (MBPT) arguments and, thus, it may give rise to an unphysical description of the system, such as a bump in the potential energy curve (PEC) along bond dissociation coordinates, especially when restricted Hartree--Fock (HF) references are used. Various non-iterative triples corrections aim to eliminate this unphysical behavior while faithfully recovering a CCSDT-level description of the systems of 
{\color{black} interest. For example, there have been attempts to incorporate the left-hand CCSD eigenvector in the non-iterative correction, which is justifiable due to the non-symmetric nature of the CC similarity-transformed Hamiltonian, leading to the $\Lambda$-CCSD(T) or CCSD(T)$_\Lambda$ method.\cite{Stanton97_130,Stanton98_601,Bartlett98_5243,Bartlett98_5255,Bartlett08_044110,Bartlett08_044111,Stanton24_e2252114} This approach has been shown to be particularly accurate when considering corrections to higher-order CC expansions.\cite{Stanton24_e2252114} 

A different strategy, which discards perturbative arguments in favor of systematically-improvable approximations to the exact left-hand state, is given by the completely renormalized (CR) CC(2,3)\cite{Wloch05_224105,Gour06_2149,Kinal06_467,Piecuch07_11359,Wloch08_2128} methodology. Similar to the $\Lambda$-based CCSD(T) approach, CR-CC(2,3) has shown promise as a correction to CCSD energetics, while also avoiding the potential pitfalls of CCSD(T).}
Indeed, CR-CC(2,3) is able to correctly describe PECs where CCSD(T) overcorrelates in the dissociation limit, and it {\color{black} can outperform} CCSD(T) in reproducing the full CCSDT {\color{black} or other high-level theoretical descriptions} for many chemically interesting situations.
{\color{black} Some challenging problems to which this approach has been applied include the description of} the homolytic dissociation of alkaline earth {\color{black} dimers,\cite{Piecuch18_1350,Piecuch19_1486}}  singlet--triplet gaps in biradical
{\color{black} species,\cite{Piecuch07_11359} relative energies of various coordinated Cu$_2$O$_2$ isomers,\cite{Gagliardi06_1991} the bromination of benzene,\cite{Pliego21_113171} and bicyclobutane to butadiene isomerization pathways.\cite{Piecuch07_734}}
The CR-CC(2,3) method is part of a more general non-iterative correction scheme called CC($P$;$Q$) introduced in Refs.~\citenum{Piecuch12_180,Piecuch12_144104,Piecuch12_4968}, which accounts for arbitrary truncation of the cluster operator. While not the focus of the present work, it is worth noting that these approaches can be extended to treat electronically excited  states\cite{Piecuch05_214107,Gour06_2149,Wloch09_3268,Wloch11_1647} via the equation-of-motion (EOM)\cite{Emrich81_379,Bartlett93_7029} framework.

Aside from high-order correlation effects, accurate descriptions of spin-orbit coupling, core-excited states, and molecules containing heavy elements require the consideration of relativistic effects. The exact two-component (X2C)\cite{Dyall97_9618,Dyall98_4201,Enevoldsen99_10000,Dyall01_9136,Cremer02_259,Liu05_241102,Peng06_044102,Cheng07_104106,Saue07_064102,Peng09_031104,Liu10_1679,Liu12_154114,Reiher13_184105,Li16_3711,Li16_104107,Repisky16_5823,Li17_2591,Cheng21_e1536,Li22_2947,Li22_2983,Li22_5011} approach has gained popularity owing to its ability to downfold relativistic effects from the full four-component (4c) Hamiltonian into a more computationally manageable two-component (2c) framework. 
Furthermore, while 4c relativistic calculations require the use of uncontracted basis sets to avoid variational collapse or prolapse,\cite{FaegriJr01_252,Mochizuki03_399,Mochizuki03_40,Haiduke22_1901} X2C benefits from having the ability to use a re-contracted basis set after the 4c$\rightarrow$2c transformation, further reducing the computational cost.\cite{Li24_3408} For quantum chemistry applications, it is commonplace to use the X2C framework at the mean-field level and then employ the no-virtual-pair approximation\cite{Peierls97_552,Mittleman71_893,Sucher80_348,Sucher84_703,Hess86_3742} in post-HF calculations such that one only has to account for electron-electron correlation effects, without noticeable loss of accuracy. Indeed, the X2C Hamiltonian has been successfully applied to heavy elements, core-excited phenomena, and core or valence spectral splitting when combined with CC approaches or other methods that include electronic correlations.\cite{Ilias09_124116,Visscher14_041107,Gomes18_174113,Cheng18_034106,Cheng18_144108,Cheng19_074102,Li19_6617,Visscher21_5509,Kohn21_124101,Saue22_114106,Cheng22_151101,Li22_2171,Li22_2947,Li22_2983,Li22_5011,Stanton24_e2252114}
{\color{black}We also note that a number of groups have previously incorporated relativistic effects in CC calculations through approaches other than X2C, including full 4c calculations\cite{Bartlett90_241,Kaldor92_8455,Kaldor93_137,Ishikawa94_1724,Dyall96_8769,Kaldor96_405,Hess98_3409,Kaldor01_9720,Olsen07_347,Visscher10_234109} and 2c calculations using the Douglas-Kroll-Hess transformation.\cite{Hess94_1,Hess98_3409,Kaldor00_1809,Schimmelpfennig01_9667,Fan07_024104,Visscher10_234109,Nakajima17_827}}

Focusing on X2C-based single-reference CC calculations beyond the CCSD level, several studies have employed spin-free and spin-orbit coupling versions of X2C-CCSD(T) (see, {\em e.g.}, Refs.~\citenum{Cheng18_034106,Cheng18_144108,Kohn21_124101,Cheng22_151101}), 
but we are not aware of any that combined the X2C relativistic treatment with CR-CC(2,3). In this paper, we present our implementation of CCSD(T) and CR-CC(2,3) within the mean-field X2C framework, with spin-orbit coupling effects. The coinage metal dimers Cu$_2$, Ag$_2$, and Au$_2$ serve as test systems for our CCSD, CCSD(T), and CR-CC(2,3) calculations. Prior experimental and theoretical investigations of these systems have yielded reliable energetics and spectroscopic {\color{black}parameters (see Refs.~\citenum{Travis65_171,Pelissier81_775,Beckmann82_945,Valentini86_6560,Morse86_1049,Gudeman91_39,Bernath92_468,Hirao01_4463,Peterson05_064107,Puzzarini05_283,Liu05_63,Wang17_88,Radi17_214308,Radi19_3270,Radi20_244305,Candido21_9832,Lindkvist55_385,Vujisic91_516,Langridge-Smith91_415,Demtroder93_2699,Balasubramanian93_7092,Fantucci99_3876,Antic-Jovanovic08_111,Antic-Jovanovic11_786,Barrow67_39,Hackett90_310,Pitzer79_293,McLean79_288,Boyd89_1762,Lineberger90_6987,Balasubramanian90_280,Morse94_248,Schwerdtfeger99_9457,Schwerdtfeger00_9356,Kaldor00_1809} for selected examples)} and it has been shown, especially for the heavier dimers, that relativistic effects can significantly change the description of the PECs. 
{\color{black}Previous CC calculations on these systems\cite{Hirao01_4463,Peterson05_064107,Puzzarini05_283,Wang17_88,Radi20_244305,Fantucci99_3876,Schwerdtfeger99_9457,Schwerdtfeger00_9356,Kaldor00_1809} have provided accurate results, but these studies relied on the use of effective core potentials (ECPs) or the frozen core approximation and thus correlated only the valence and semi-core electrons. In some cases\cite{Fantucci99_3876,Schwerdtfeger99_9457,Schwerdtfeger00_9356}, the relativistic treatment was limited to spin-free effects that were incorporated through the ECP, while a few other studies\cite{Puzzarini05_283,Wang17_88} added spin-free and spin-orbit effects on top of the ECP treatment. Other calculations\cite{Kaldor00_1809,Peterson05_064107,Hirao01_4463,Radi20_244305} employed the frozen-core approximation in conjunction with spin-free or spin-free plus spin-orbit coupling effects for the valence electrons.}
{\color{black}T}his work {\color{black}presents} all-electron  CC calculations on these systems {\color{black} with relativistic effects captured via the X2C framework. W}e also study the effects of basis set truncation level and contraction scheme on the results of all-electron X2C-CC computations, highlighting the deficiencies of relativistic basis sets not specifically optimized for all-electron calculations with spin-orbit coupling, which has not received much attention in the previous works relying on frozen-core approximations or using ECPs.

The remaining parts of this paper are organized as follows. In Section \ref{SEC:THEORY}, we provide a short summary of the X2C formalism, CC theory, and the derivation of CR-CC(2,3) and its relationship with the CCSD(T) approach. We provide the relevant details of our computational protocols for obtaining the PECs of the coinage metal dimers in Section \ref{SEC:COMPUTATIONAL_DETAILS}. We discuss our findings in Section \ref{SEC:RESULTS}, focusing on the impact of triples correlation effects, the contrast between relativistic and {\color{black} non-}relativistic calculations, and the effects of basis set size and contraction scheme, and we provide a concluding summary in Section \ref{SEC:CONCLUSIONS}.

\section{Theory}

\label{SEC:THEORY}

\subsection{Exact two-component transformation}
The details of the X2C transformation employed in this work have been described elsewhere (see, Ref.~\citenum{Reiher13_184105}, for example), so we provide here a summary of the relevant elements. Throughout this work, we are concerned with the electronic Hamiltonian, which in second quantization can be expressed as
\begin{equation}
    \label{eqn:electronic_hamiltonian}
    \hat{H} = h_p^q \hat{a}^p \hat{a}_q
            + \tfrac{1}{4} g_{pq}^{rs} \hat{a}^p \hat{a}^q \hat{a}_s \hat{a}_r.
\end{equation}
In Eq.~(\ref{eqn:electronic_hamiltonian}), $h_p^q$ and $g_{pq}^{rs}$ are (antisymmetrized) matrix elements of the one- and two-electron parts of the Hamiltonian, and $\hat{a}_p$ and $\hat{a}^p \equiv \hat{a}_p^\dagger$ are the fermionic annihilation and creation operator associated with the spin-orbital label $p$. 
{\color{black} The indices $p,q,\ldots$ refer to general spinor labels.}
In this paper, we use the Einstein convention where repeated lower and upper indices imply summation. 
{\color{black} In the non-relativistic limit, $h_p^q$ contains the kinetic energy of the electrons and nuclei-electron attraction, whereas $g_{pq}^{rs}$ describe the Coulombic repulsion between electron pairs.}
In the X2C framework, we begin with the 4c relativistic Dirac Hamiltonian for the {\color{black} electrons,
which modifies the one-body part to include one-electron scalar (spin-free) relativistic effects and spin-orbit coupling effects on top of the non-relativistic description,}
with a restricted-kinetic-balanced condition,\cite{Faegri07_book,Reiher15_book,Liu17_handbook}
\begin{equation}
    \label{eqn:4c_hamiltonian}
    \hat{H}^\mathrm{4c} =
    \begin{pmatrix}
        \mathbf{V}\mathbf{I} & \mathbf{T}\mathbf{I} \\
        \mathbf{T}\mathbf{I} & \mathbf{W}-\mathbf{T}\mathbf{I}
    \end{pmatrix}.
\end{equation}
In Eq.~(\ref{eqn:4c_hamiltonian}), $\mathbf{V}$ is the {\color{black} matrix representation of the} scalar potential, $\mathbf{I}$ is the $2\times2$ identity matrix, $\mathbf{T}$ is the {\color{black}matrix representation of the} kinetic energy, and
\begin{equation}
    \label{eqn:w}
    \mathbf{W} = \frac{1}{4m^2c^2}({\color{black}\bm{\sigma}}\cdot\mathbf{p})\mathbf{V}({\color{black}\bm{\sigma}}\cdot\mathbf{p}),
\end{equation}
where $m$ is the mass of an electron, $c$ is the speed of light, {\color{black} $\bm{\sigma}$} is the vector of Pauli matrices, and $\mathbf{p}$ is the linear momentum operator. The eigenstates of this Hamiltonian can be expressed as
\begin{equation}
    \label{eqn:large_small_components}
    \Psi^\mathrm{4c} = 
    \begin{pmatrix}
        \Psi_L \\
        \Psi_S
    \end{pmatrix},
\end{equation}
in which $\Psi_L$ and $\Psi_S$ are the large and small components, respectively, that can be further broken down into their $\alpha$ and $\beta$ spin components. Solving this eigenvalue problem yields positive- and negative-energy states, the former of which are of interest in the electronic structure calculations. To isolate the positive-energy states, one may perform a unitary transformation
\begin{equation}
    \label{eqn:x2c_unitary}
    \mathbf{U}^\dagger \hat{H}^\mathrm{4c} \mathbf{U} = 
    \begin{pmatrix}
        \hat{H}^+  & \mathbf{0} \\
        \mathbf{0} & \hat{H}^-
    \end{pmatrix},\quad
    \mathbf{U}^\dagger \Psi^\mathrm{4c} = 
    \begin{pmatrix}
        \Psi^\mathrm{2c} \\
        \mathbf{0}
    \end{pmatrix},
\end{equation}
where $\hat{H}^\pm$ describe the positive- and negative-energy states and $\Psi^\mathrm{2c}$ are the 2c eigenstates of these Hamiltonians. In this work, the 4c$\rightarrow$2c transformation described in Eq.~(\ref{eqn:x2c_unitary}) is applied to the one-electron part of the Hamiltonian, while the two-electron part is treated {\color{black} non-}relativistically. In this approximation, which is usually denoted as 1eX2C, we applied an empirical screened-nuclei spin-orbit (SNSO)\cite{Boettger00_7809,Li23_5785} scaling factor to the one-body integrals to mimic {\color{black}two-electron spin-orbit coupling effects. An alternative approach not pursued in this work is to incorporate two-electron spin-orbit effects at the mean-field level, which has been done in the context of CC theory using the atomic mean-field or molecular mean-field approaches.\cite{Ilias09_124116,Visscher14_041107,Cheng18_034106,Gomes18_174113,Cheng19_074102,Visscher21_5509} In either case,} the X2C transformation is done at the mean-field ({\em i.e.}, HF) level and we employ the no-virtual-pair approximation in {\color{black}subsequent} correlated calculations. {\color{black} We refer the reader to  Ref.~\citenum{Li23_5785} and the references cited therein for an explanation of how the two-electon spin-orbit contributions arise in the Dirac--Coulomb--Breit Hamiltonian, the rationale behind the SNSO approach, and more detailed descriptions of the atomic mean field and molecular mean field approaches for incorporating two-electron spin-orbit interactions within the X2C framework.}

\subsection{Coupled-cluster theory}

The ground-state CC wave function for an $N$-electron system is given by the exponential {\em ansatz}
\begin{equation}
    \label{eqn:cc_wfn}
    \ket{\Psi} = e^{\hat{T}} \ket{\Phi},
\end{equation}
where $\hat{T}$ is the cluster operator, and $\ket{\Phi}$ is a reference Slater determinant (here, a HF determinant). The cluster operator is expanded by excitation order as
\begin{equation}
    \label{eqn:cluster_operator}
    \hat{T} = \sum_{n=1}^{N} \hat{T}_n,\quad
    \hat{T}_n = t_{a_1\ldots a_n}^{i_1\ldots i_n} E_{i_1\ldots i_n}^{a_1\ldots a_n},
\end{equation}
where $t_{a_1\ldots a_n}^{i_1\ldots i_n}$ is a cluster amplitude and $E_{i_1\ldots i_n}^{a_1\ldots a_n} = \prod_k^n \hat{a}^{a_k} \hat{a}_{i_k}$ is the usual $n$-body particle--hole excitation operator, which generates the manifold of excited determinants $\ket*{\Phi_{i_1\ldots i_n}^{a_1\ldots a_n}} = E_{i_1\ldots i_n}^{a_1\ldots a_n} \ket{\Phi}$. We use the indices $i_1,i_2,\ldots$ ($i,j,\ldots$) and $a_1,a_2,\ldots$ ($a,b,\ldots$) to designate the spin orbitals that are occupied and unoccupied, respectively, in the reference determinant. The cluster amplitudes are obtained by solving a system of non-linear, energy-independent equations
\begin{equation}
    \label{eqn:cc_project}
    \mel*{\Phi_{i_1\ldots i_n}^{a_1\ldots a_n}}{\bar{H}}{\Phi} = 0\;
    \forall\; \ket*{\Phi_{i_1\ldots i_n}^{a_1\ldots a_n}},\;n = 1,\ldots,N,
\end{equation}
in which $\bar{H} = e^{-\hat{T}}\hat{H}e^{\hat{T}}$ is the similarity-transformed Hamiltonian, and the ground-state energy is computed as the expectation value
\begin{equation}
    \label{eqn:cc_energy}
    E^\mathrm{(CC)} = \mel*{\Phi}{\bar{H}}{\Phi}.
\end{equation}

Due to the non-hermiticity of $\bar{H}$, if properties other than energies are desired, one also needs to solve the left-hand CC problem. The left-hand CC wave function is parameterized as
\begin{equation}
    \label{eqn:left_cc_wfn}
    \bra{\tilde{\Psi}} = \bra{\Phi} (\mathbf{1}+\hat{\Lambda}) e^{-\hat{T}},
\end{equation}
where $\hat{\Lambda}$ is the many-body de-excitation operator
\begin{equation}
    \label{eqn:lambda_operator}
    \hat{\Lambda} = \sum_{n=1}^{N} \hat{\Lambda}_n,\quad
    \hat{\Lambda}_n = \lambda_{i_1\ldots i_n}^{a_1\ldots a_n} (E_{i_1\ldots i_n}^{a_1\ldots a_n})^\dagger.
\end{equation}
One obtains the $\lambda_{i_1\ldots i_n}^{a_1\ldots a_n}$ amplitudes by solving a linear system of equations
\begin{equation} 
    \label{eqn:left_cc_equations}
    \mel*{\Phi}{(\mathbf{1}+\hat{\Lambda})(\bar{H}-\mathbf{1}E^\mathrm{(CC)})}{\Phi_{i_1\ldots i_n}^{a_1\ldots a_n}} = 0~~~~~~\forall\; \ket*{\Phi_{i_1\ldots i_n}^{a_1\ldots a_n}},\;n = 1,\ldots,N
\end{equation}

It is worth mentioning that the single-reference CC formalism as described above is equivalent to the full CI methodology. However, in practice, the many-body expansion of $\hat{T}$ and $\hat{\Lambda}$, along with the corresponding projection spaces used in Eqs.~(\ref{eqn:cc_project}) and (\ref{eqn:left_cc_equations}), are truncated at a computationally tractable level (much lower than $N$), giving rise to the usual CCSD, CCSDT, {\em etc.}~hierarchy. As mentioned in Section \ref{SEC:INTRODUCTION}, it is well known that CCSD is often not sufficient to produce quantitatively accurate results and, thus, one of the main objectives in the field is the incorporation of correlation effects due to higher--than--doubly excited clusters without incurring significant computational costs.

\subsection{Non-iterative triples corrections to CCSD energetics}
In this work, we focus on the CR-CC(2,3) correction to CCSD energetics. Let us recall that the CR-CC approaches arise from the biorthogonal moment expansion derived via the method-of-moments of CC equations (MMCC)\cite{Kowalski00_1,Piecuch00_18,Piecuch00_5644,Mcguire02_527,Musial04_349,Piecuch05_074107,Piecuch05_2191,Piecuch07_11359,Wloch09_3268,Piecuch12_180} framework, which avoids using MBPT analysis in incorporating the higher-order corrections to lower-order CC energies. The derivation of CR-CC(2,3) working equations from the asymmetric energy expression has been discussed in great detail in the literature (see, {\em e.g.}, Refs.~\citenum{Wloch05_224105,Kinal06_467,Gour06_2149,Piecuch07_11359}), and we only provide a summary in this subsection. In a CR-CC(2,3) calculation, one begins by solving the right- and left-hand CCSD problems by setting $\hat{T} = \hat{T}_1 + \hat{T}_2$ and $\hat{\Lambda} = \hat{\Lambda}_1 + \hat{\Lambda}_2$ and solving Eqs.~(\ref{eqn:cc_project}) and (\ref{eqn:left_cc_equations}) in the space of singly and doubly substituted determinants. Subsequently, the CCSD energy is corrected using the CR-CC(2,3) moment expansion
\begin{equation}
    \label{eqn:crcc_correction}
    \delta(2,3) = \tfrac{1}{36} \ell_{ijk}^{abc}(2) \mathfrak{M}_{abc}^{ijk}(2).
\end{equation}
The moments $\mathfrak{M}_{abc}^{ijk}(2)$ are defined as the projections of the Schr{\"o}dinger equation containing the CCSD wave function onto triply excited determinants,
\begin{equation}
    \label{eqn:crcc_m3}
    \mathfrak{M}_{abc}^{ijk}(2) = \mel*{\Phi_{ijk}^{abc}}{\bar{H}^\mathrm{(CCSD)}}{\Phi},
\end{equation}
where $\bar{H}^\mathrm{(CCSD)} = e^{- \hat{T}_1 - \hat{T}_2} \hat{H} e^{\hat{T}_1 + \hat{T}_2}$. Note that the triples moments in Eq.~(\ref{eqn:crcc_m3}) are non-zero because these projections are not part of the CCSD amplitude equations. The $\ell_{ijk}^{abc}(2)$ amplitudes entering Eq.~(\ref{eqn:crcc_correction}) are defined as
\begin{equation}
    \label{eqn:crcc_l3}
    \ell_{ijk}^{abc}(2) = \mel*{\Phi}{(\hat{\Lambda}_1+\hat{\Lambda}_2)\bar{H}^\mathrm{(CCSD)}}{\Phi_{ijk}^{abc}}/D_{ijk}^{abc},
\end{equation}
where we have introduced a quasi-perturbative denominator $D_{ijk}^{abc}$, which is given by the Epstein--Nesbet-like expression
\begin{equation}
    \label{eqn:crcc_denom}
    D_{ijk}^{abc} = E^\mathrm{(CCSD)} - \mel*{\Phi_{ijk}^{abc}}{\bar{H}^\mathrm{(CCSD)}}{\Phi_{ijk}^{abc}}.
\end{equation}

The above set of equations describe the most complete variant of CR-CC(2,3), usually denoted as CR-CC(2,3)$_\mathrm{D}$, which includes up to the three-body component of $\bar{H}^\mathrm{(CCSD)}$ in Eq.~(\ref{eqn:crcc_denom}). One could introduce a simplification by including only the one-body component or the one- and two-body parts of $\bar{H}^\mathrm{(CCSD)}$ in the denominator, resulting in the CR-CC(2,3)$_\mathrm{B}$ and CR-CC(2,3)$_\mathrm{C}$ variants, respectively, or even go further and replace the Epstein--Nesbet-like expression in Eq.~(\ref{eqn:crcc_denom}) by its M{\o}ller--Plesset variant involving orbital energy differences, to obtain the simplest CR-CC(2,3)$_\mathrm{A}$ approach (which is equivalent to CCSD(2)$_\mathrm{T}$ of Ref.~\citenum{Piecuch04_12197}). As has been discussed and demonstrated extensively,\cite{Wloch05_224105,Gour06_2149,Kinal06_467,Piecuch07_11359,Wloch08_2128} the most complete D variant usually outperforms its A--C approximate versions. However, due to the use of Epstein--Nesbet denominator, CR-CC(2,3)$_\mathrm{D}$ is not invariant with respect to rotations of degenerate orbitals.\cite{Piecuch07_11359} In fact, only the A and B versions of CR-CC(2,3) are orbital invariant, because these approaches use only the (effective) one-body part of the Hamiltonian in the denominator. One could avoid this issue by using the $\ell_{ijk}^{abc}(2)$ amplitudes involving degenerate orbitals that are obtained through solving a linear system of equation in the appropriate subspace (see, {\em e.g.}, Ref.~\citenum{Varandas07_63} for more details). The arbitrary mixing of the spatial parts of degenerate orbitals can be cleaned up through the use of point group or double point group symmetry, but issues may still persist regarding $\alpha$ and $\beta$ spin mixing that can occur in relativistic calculations with spin-orbit coupling effects. Nevertheless, in this work, we opt to implement all variants of CR-CC(2,3) following the above description to ensure reproducibility with existing CR-CC(2,3) implementations, assuming that the exact same set of molecular orbitals are used.

Before moving on, it is worth mentioning that the CCSD(T) methodology can be understood from the lens of CR-CC(2,3) by making several modifications. Specifically, one uses a simpler form of Eq.~(\ref{eqn:crcc_correction}), in which one makes the approximations
\begin{equation}
    \label{eqn:ccsdpt_m3}
    \mathfrak{M}_{abc}^{ijk}(2) \approx
        \mel*{\Phi_{ijk}^{abc}}{(\hat{V}_N\hat{T}_2)_C}{\Phi}
\end{equation}
and
\begin{equation}
    \label{eqn:ccsdpt_l3}
    \ell_{ijk}^{abc}(2) \approx
        \mel*{\Phi}{(\hat{V}_N\hat{T}_1)_\mathrm{DC}^\dagger + (\hat{V}_N\hat{T}_2)_\mathrm{C}^\dagger}{\Phi_{ijk}^{abc}}/D_{ijk}^{abc}.
\end{equation}
Here, $\hat{V}_N$ is the normal-ordered two-body part of the Hamiltonian, the subscripts C and DC refer to connected and disconnected diagrams, respectively, and we employ the M{\o}ller--Plesset form of $D_{ijk}^{abc}$ in Eq.~(\ref{eqn:ccsdpt_l3}).

\section{Algorithmic and Computational Details}

\label{SEC:COMPUTATIONAL_DETAILS}

{\color{black}
\subsection{Correlated 1eX2C-based calculations}

This subsection provides some technical details of our 1eX2C-based CC codes, which are implemented in a development version of the Chronus Quantum software package.\cite{Li20_e1436} Let us examine the 2c Hamiltonian in Eq.~(\ref{eqn:electronic_hamiltonian}), which contains both one- and two-body pieces. In the 1eX2C scheme, the one-body part of this 2c Hamiltonian is obtained via the following protocol. One first expands the relativistic one-body Hamiltonian ({\em i.e.}, the Dirac Hamiltonian in Eq.~\ref{eqn:4c_hamiltonian}) in the 4c, uncontracted atomic basis, including both one-electron scalar relativity and spin-orbit coupling. This Hamiltonian is then brought to block-diagonal form via the exact-two-component (X2C) unitary transformation approach (Eq.~\ref{eqn:x2c_unitary}; see, {\em e.g.}, Refs.~\citenum{Reiher13_184105} for further details regarding how this unitary transformation matrix is determined).  After this transformation, we isolate the 2c positive-energy ({\em i.e.}, electronic) part of the Hamiltonian and recontract the basis using the basis set's original contraction coefficients.\cite{Li24_3408} 
To account for the two-electron spin-orbit coupling, an empirical screened-nuclear spin-orbit factor\cite{Boettger00_7809,Li23_5785} is utilized to scale the one-electron spin-orbit term.
In the two-body part of the 2c Hamiltonian, the untransformed Coulomb operator is used. This 2c Hamiltonian is then used within a standard HF procedure that supports complex arithmetic. Additional technical details of the 1eX2C-HF implementation in Chronus Quantum can be found in Ref.~\citenum{Li20_e1436} and the references therein. 

Subsequent correlated calculations are carried out using the basis of complex molecular spinors obtained via the 1eX2C-HF procedure. As such, a 1eX2C-CCSD implementation, for example, resembles a standard, non-relativistic CCSD implementation except that the cluster amplitudes and integrals are complex-valued and lack any spin symmetry ({\em i.e.}, they must be expanded in the full 2c spinor basis). Similar generalizations must be made for the left-hand CCSD, CCSD(T), and CR-CC(2,3) implementations. The 1eX2C-CCSD and left-hand 1eX2C-CCSD residual equations were implemented in Chronus Quantum using the TiledArray tensor algebra library.\cite{TiledArray} In this implementation, the complex-valued two-electron integrals are stored in memory; storage requirements are met by executing the code in a distributed memory environment. The 1eX2C-CCSD(T) and 1eX2C-CR-CC(2,3) equations were also implemented using TiledArray, but, as discussed in the next subsection, our codes make use of a loop-based algorithm to maintain manageable storage requirements. 

\subsection{Considerations for triples corrections to 1eX2C-CCSD}

The (T) and CR-CC(2,3) triples corrections to 1eX2C-CCSD are implemented in a loop-based algorithm similar to that of Refs.~\citenum{Komornicki91_462,Musial02_71,Piecuch05_214107,Wloch09_3268}, but without the permutational symmetry that can be exploited in non-relativistic calculations on closed-shell systems.\cite{Komornicki91_462} Within the loop structure itself, one needs to construct several $n_o^3$/$n_u^3$-sized objects per batch by looping over unoccupied/occupied indices. Using this strategy, our implementation of CCSD(T) requires the aggregate storage of two-body integral slices $v_{mn}^{ef}$ ($n_o^2 n_u^2$), $v_{mb}^{ij}$ ($n_o^3 n_u$), and $v_{ab}^{ej}$ ($n_o n_u^3$), the $\hat{T}_1$ ($n_o n_u$) and $\hat{T}_2$ ($n_o^2 n_u^2$) amplitudes, and the orbital energies ($n_o + n_u$). In terms of floating point operations, the most expensive contraction in CCSD(T) scales as $\mathcal{O}(n_o^3 n_u^4)$, involving the $v_{ab}^{ej} t_{ec}^{ik}$-type contractions, on top of the iterative $\mathcal{O}(n_o^2 n_u^4)$ CCSD step.

In the CR-CC(2,3) triples correction, one needs to account for the aggregate storage of the integral slices $v_{mn}^{ef}$ ($n_o^2 n_u^2$), $v_{mb}^{ij}$ ($n_o^3 n_u$), and $v_{ab}^{ej}$ ($n_o n_u^3$), the $\bar{H}$ intermediates $\bar{h}_{m}^{e}$ ($n_o n_u$), $\bar{h}_{mb}^{ij}$ ($n_o^3 n_u$), and $\bar{h}_{ab}^{ej}$ ($n_o n_u^3$), the $\hat{T}_1$/$\hat{\Lambda}_1$ ($2\times n_o n_u$) and $\hat{T}_2$/$\hat{\Lambda}_2$ ($2\times n_o^2 n_u^2$) amplitudes, and the \textit{diagonal} 1-body ($n_o + n_u$), 2-body ($n_o^2 + n_o n_u + n_u^2$), and 3-body ($n_o^2 n_u + n_o n_u^2$) $\bar{H}$ elements for the Epstein--Nesbet denominator or the orbital energies ($n_o + n_u$) for the M{\o}ller--Plesset variant.
Similar to the CCSD(T) case, the most expensive contractions in CR-CC(2,3) scale as $\mathcal{O}(n_o^3 n_u^4)$, but there are two such contractions in CR-CC(2,3), of the forms $\bar{h}_{ab}^{ej} t_{ec}^{ik}$ and $\lambda_{ik}^{ec} \bar{h}_{ej}^{ab}$. Given that one of these contractions involves the de-excitation amplitudes from the left-hand CCSD problem [Eq.~(\ref{eqn:crcc_l3})], in contrast to the simpler conjugate form found in CCSD(T) [Eq.~\ref{eqn:ccsdpt_l3})], the overall CR-CC(2,3) procedure carries roughly twice the floating-point cost of CCSD(T), in both the iterative (CCSD / left-hand CCSD) and non-iterative (triples corrections) parts of the algorithm.
We refer the reader to, {\em e.g.}, Refs.~\citenum{Musial02_71,Piecuch05_214107,Piecuch07_11359,Wloch09_3268}, for further details on CR-CC implementations, including recursive, on-the-fly construction of the $\bar{H}$ intermediates to reduce storage requirements, as well as Refs.~\citenum{Valeev19_e25894,Valeev21_1203} for thorough reviews of parallel implementation of CCSD(T)-type methods.

\subsection{Calculation details}
}

Our objectives in this paper are to demonstrate the improvements that CCSD(T) and CR-CC(2,3) deliver on top of CCSD energetics in the context of relativistic calculations,
as well as to highlight the sensitivity of all-electron correlated calculations to the choice of basis set contraction scheme and size.
To that end, we computed the ground-state PECs of Cu$_2$, Ag$_2$, and Au$_2$ using the CCSD, CCSD(T), and CR-CC(2,3) (variants A--D) approaches. In particular, we focus on the dissociation energy ($D_\mathrm{e}$), equilibrium bond distance ($R_\mathrm{e}$), and fundamental vibrational frequency ($\omega_\mathrm{e}$) characterizing the PECs obtained using the ANO-RCC-VDZP basis set.\cite{Widmark05_6575} In addition to the different levels of electronic correlation treatment, we also examine how relativistic effects modify the PECs by comparing the results based on generalized HF (GHF), spin-free 1eX2C-HF (SF-1eX2C-HF), and 1eX2C-HF reference functions.
In the case of the 1eX2C-HF calculations, we utilized the row-dependent SNSO scaling factors of Ref.~\citenum{Li23_5785} that were parameterized using the full 4c Dirac--Coulomb--Breit Hamiltonian. 

Our CC implementations are based on the Kramers-unrestricted formalism, and, thus, are separable at the dissociation asymptote. Nevertheless, to provide a more complete picture of how the different triples corrections behave, for each dimer we computed the potentials that follow both the Kramers-unrestricted/``spin-broken'' [{\em i.e.}, $\expval*{\hat{S}^2} = 1$ and $E(\mathrm{X}_2) = 2E(\mathrm{X})$ at the dissociation 
{\color{black} limit, similar to the non-relativistic unrestricted HF scheme]}
as well as Kramers-restricted/``spin-pure'' ({\em i.e.}, $\expval*{\hat{S}^2} = 0$ throughout the entire 
{\color{black} curve, analogous to the non-relativistic restricted HF scheme)}
reference PECs. The potential curves are computed at the points on the grid defined in Table \ref{tab:dist_grid}. Subsequently, we fit each PEC at the lowest-energy point in the grid and the {\color{black} next two points in both direction ({\em i.e.}, 5 points in total) }
to a {\color{black} 4th-order} polynomial, and use the resulting information to obtain our $D_\mathrm{e}$, $R_\mathrm{e}$, and $\omega_\mathrm{e}$ estimates. {\color{black} The electronic energies at the grid points shown in Table \ref{tab:dist_grid}, the corresponding fit parameters, and additional spectroscopic parameters, including anharmonicity constant ($\omega_\mathrm{e}\chi_\mathrm{e}$) and the dissociation energy relative to the ground vibrational state ($D_0$), are provided in the Supporting Information.}

\begin{table}[!htbp]
    \caption{
        \label{tab:dist_grid}
        Grids of internuclear distances (in \AA) employed in the PEC calculations of Cu$_2$, Ag$_2$, and Au$_2$.}
    \centering
    \begin{tabular*}{\columnwidth}{@{\extracolsep{\fill}}cccc}
        \hline \hline
        Step size & Cu$_2$     & Ag$_2$     & Au$_2$ \\
        \hline
        0.10 & 1.70--2.50 & 2.00--2.90 & {\color{black}1.60}--2.90 \\
        0.50 & 3.00--5.00 & 3.00--5.00 & 3.00--5.00 \\
        1.00 & 6.00--8.00 & 6.00--8.00 & 6.00--8.00 \\
        \hline \hline
    \end{tabular*}
\end{table}

In addition to the PEC studies using the ANO-RCC-VDZP basis set, we investigate the convergence of our CC calculations with respect to the basis set size by using the ANO-RCC-VTZP, ANO-RCC-VQZP, and ANO-RCC sets. In the case of gold dimer, we also performed additional computations using the full ANO-RCC set with an extra $i$-type primitive (exponent = 15.1665360) obtained from the ``dyall.ae4z'' basis set.\cite{Dyall12_1217,Dyall23_} We also compared the convergence of the all-electron CC calculations using the ANO-RCC series with the x2c-SVPall-2c, x2c-TZVPall-2c, x2c-TZVPPall-2c, and x2c-QZVPPall-2c basis sets,\cite{Weigend17_3696,Weigend20_5658} which were optimized in all-electron 1eX2C-HF calculations including spin-orbit coupling effects. All of these calculations were performed at the grid point closest to the equilibrium bond distance found in the 1eX2C-CCSD/ANO-RCC-VDZP calculations for each of the dimers.

All of the electronic structure calculations reported in this work were performed using a development branch of the Chronus Quantum software package.\cite{Li20_e1436} All basis sets were extracted from the Basis Set Exchange.\cite{Feller96_1571,Windus07_1045,Windus19_4814} The CCSD(T) and CR-CC(2,3) working equations were derived with the help of the \texttt{p$^\dagger$q} automated code generator\cite{DePrince21_e1954709} and implemented using the TiledArray\cite{TiledArray} high-performance tensor arithmetic framework. The resulting code was benchmarked numerically against the CC routines\cite{Musial02_71,Piecuch05_214107} in GAMESS version 2022 
{\color{black}
R2\cite{Gordon20_154102,Gordon23_7031} and against independent implementations of 1eX2C-CCSD and SF-1eX2C-CCSD(T) in MPQC.\cite{mpqc}
}
In the relativistic calculations, the speed of light $c = 137.035999074$ a.u.~was used. {\color{black} For reproducibility, the harmonic frequencies and anharmonicity constants were computed with the units and constants used in the \texttt{anharmonicity} function inside \textsc{Psi4}.\cite{Sherrill20_184108,psi4_anharmonicity}}

\section{Results and Discussion}

\label{SEC:RESULTS}

\subsection{Cu$_2$}
\label{SUBSECTION:Cu2}

\begin{figure*}[!htbp]
    \centering
    \includegraphics[width=\textwidth]{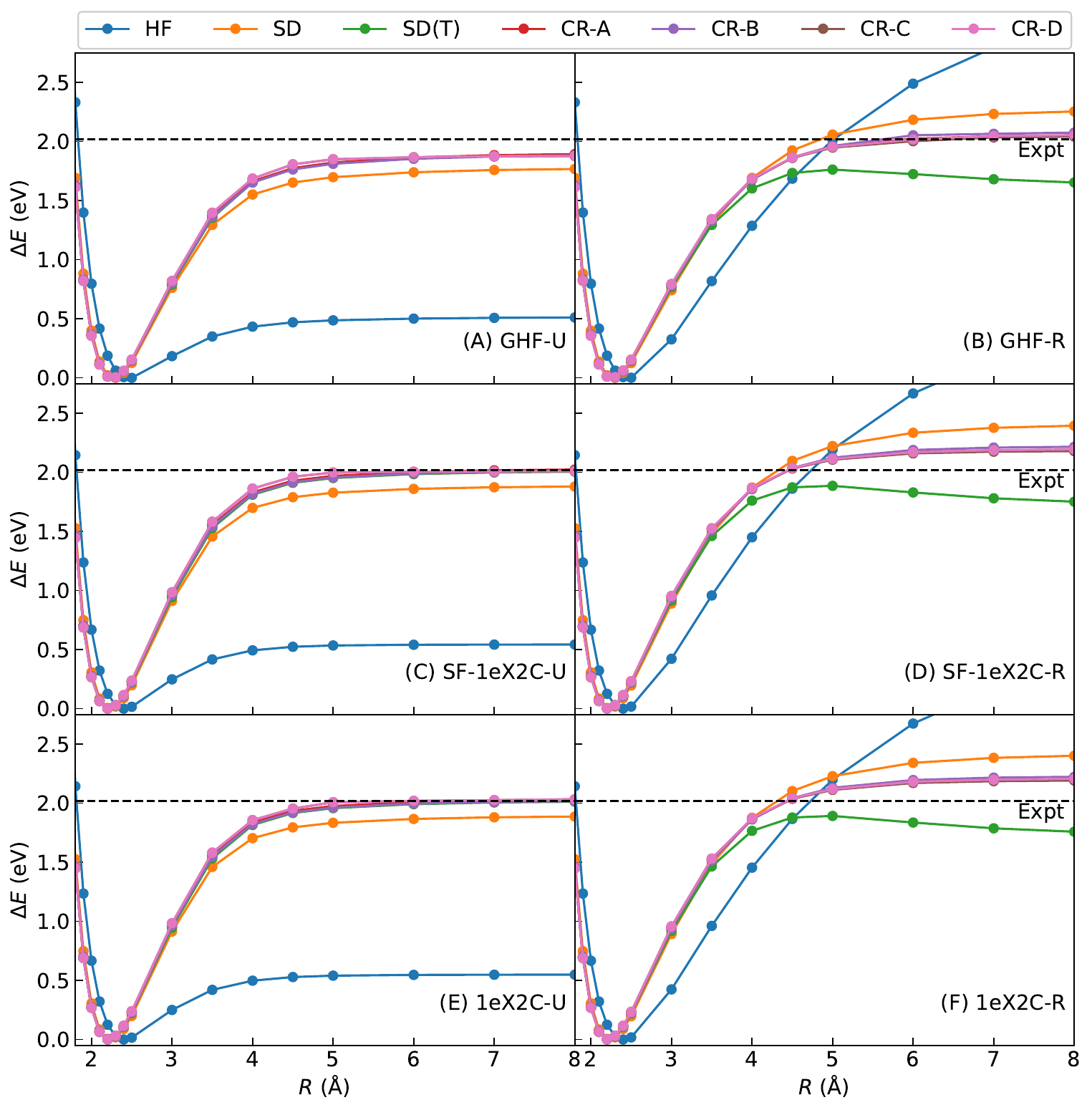}
    \caption{
        \label{fig:cu2_pec_6panels}
        The PECs of Cu$_2$ obtained in this work using the ANO-RCC-VDZP basis set. The U and R labels indicate Kramers-unrestricted and restricted reference curves, respectively. Each PEC is shifted relative to the energy at $R_\mathrm{e}$ (cf.~Table \ref{tab:cu2_vdzp}).}
\end{figure*}

\begin{sidewaystable}
\begin{minipage}{\textwidth}
    \caption{
        \label{tab:cu2_vdzp}
        Spectroscopic constants for Cu$_2$ obtained using various CC methodologies with the ANO-RCC-VDZP basis set and different levels of mean-field relativistic treatment.
        {\color{black} All values are reported as deviations from experimentally-obtained values.}}
    \centering
    {\color{black}
    \begin{tabular*}{\linewidth}{@{\extracolsep{\fill}} cccccccccc}
        \hline\hline
        \multirow{2}{*}{Method} & \multicolumn{3}{c}{$D_\mathrm{e}$ (eV)\footnotemark} & \multicolumn{3}{c}{$R_\mathrm{e}$ (\AA)} & \multicolumn{3}{c}{$\omega_\mathrm{e}$ (cm$^{-1}$)} \\
        \cline{2-4} \cline{5-7} \cline{8-10}
        & GHF & SF-1eX2C & 1eX2C & GHF & SF-1eX2C & 1eX2C & GHF & SF-1eX2C & 1eX2C \\
        \hline
        CCSD                    & $-0.25$   (0.24)  & $-0.14$   (0.37)  & $-0.13$   (0.38)  & 0.058 &   0.013  &   0.013  & $-23.2$ & $-6.2$ & $-6.2$ \\
        CCSD(T)                 & $-0.12$ ($-0.36$) & $-0.01$ ($-0.27$) &   0.00  ($-0.26$) & 0.043 & $-0.002$ & $-0.002$ & $-18.4$ & $-0.8$ & $-0.8$ \\
        CR-CC(2,3)$_\mathrm{A}$ & $-0.12$   (0.03)  &   0.01    (0.17)  &   0.01    (0.18)  & 0.041 & $-0.004$ & $-0.004$ & $-16.4$ &   1.7  &   1.7  \\
        CR-CC(2,3)$_\mathrm{B}$ & $-0.13$   (0.06)  & $-0.01$   (0.20)  &   0.00    (0.20)  & 0.043 & $-0.002$ & $-0.002$ & $-17.2$ &   0.7  &   0.7  \\
        CR-CC(2,3)$_\mathrm{C}$ & $-0.14$   (0.02)  & $-0.02$   (0.16)  &   0.00    (0.17)  & 0.038 & $-0.006$ & $-0.009$ & $-14.8$ &   1.8  &  10.9  \\
        CR-CC(2,3)$_\mathrm{D}$ & $-0.14$   (0.04)  & $-0.02$   (0.17)  &   0.00    (0.19)  & 0.038 & $-0.006$ & $-0.009$ & $-14.8$ &   1.7  &  11.1  \\
        \hline
        Experiment\footnotemark & \multicolumn{3}{c}{2.02} & \multicolumn{3}{c}{2.218} & \multicolumn{3}{c}{266.487} \\
        \hline\hline
    \end{tabular*}
    }
    \footnotetext[1]{The numbers outside and inside parentheses refer to Kramers-unrestricted and restricted dissociation energies, respectively. The latter estimates are computed as $D_\mathrm{e} = E(R = 8.00\;\mbox{\AA})-E(R = R_\mathrm{e})$.}
    \footnotetext[2]{Ref.~\citenum{Radi20_244305}.}
\end{minipage}
\end{sidewaystable}

We begin our discussion with the smallest system in our test set, the copper dimer (in particular, the $^{63}$Cu$_2$ isotopologue).
In terms of the electronic structure, we are interested in the ground $^1\Sigma_{g}^{+}$ state that dissociates into two Cu atoms, each in a $^2$S ([Ar] $3d^{10}4s^{1}$) configuration.
There have been extensive experimental and theoretical investigations of the ground electronic state of this molecule (see {\color{black}Refs.~\citenum{Travis65_171,Pelissier81_775,Beckmann82_945,Valentini86_6560,Morse86_1049,Gudeman91_39,Bernath92_468,Puzzarini05_283,Liu05_63,Wang17_88,Radi17_214308,Radi19_3270,Radi20_244305,Candido21_9832}} for selected examples). The latest experimental investigation,\cite{Radi20_244305} aided with CCSD(T) and multi-reference CI calculations, yielded a dissociation energy estimate of $D_\mathrm{e}=16,270$ cm$^{-1}$ (2.02 eV), Cu--Cu equilibrium distance of $R_\mathrm{e}=2.218$ \AA, and the harmonic frequency of $\omega_\mathrm{e}=266.487$ cm$^{-1}$, all of which can be considered to be converged results. Here, our goal is to assess different relativistic schemes and electronic correlation treatments, as compared to these existing results; our data are summarized in Fig.~\ref{fig:cu2_pec_6panels} and Table \ref{tab:cu2_vdzp}.

We begin by considering the PECs shown in Fig.~\ref{fig:cu2_pec_6panels} and the corresponding $D_\mathrm{e}$ values in Table \ref{tab:cu2_vdzp}. From Fig.~\ref{fig:cu2_pec_6panels}, we can make a few interesting observations regarding the different relativistic and electron correlation treatments, as well as whether the PECs follow the Kramers-unrestricted or restricted solutions.  First of all, the HF PECs significantly underbind and overbind the Cu$_2$ dimer in the Kramers-unrestricted and restricted cases, respectively, regardless of the specific relativistic treatment, which is not surprising.
In general, neither spin-free nor spin-orbit coupling relativistic effects have much impact on the Kramers-unrestricted HF potential wells, changing the dissociation energies by no more than 0.03 eV. These effects are somewhat more significant in the Kramers-restricted case, deepening the HF well by about 0.2 eV compared to the {\color{black} non-}relativistic case.

The differences between PECs calculated using correlated approaches are more noticeable. Beginning with CCSD, in the non-relativistic regime [Fig.~\ref{fig:cu2_pec_6panels}(A) and (B)], we observe a significant improvement over the underlying HF PECs, with dissociation asymptotes that are much closer to the experimentally determined values. However, Kramers-unrestricted GHF-CCSD still underbinds the Cu$_2$ molecule by about {\color{black} 0.25} eV, whereas its restricted counterpart overbinds Cu$_2$ by {\color{black} about the same magnitude}. 
The SF-1eX2C and 1eX2C relativistic frameworks offer a slight improvement in the Kramers-unrestricted case, reducing the CCSD error to {\color{black} less than $-0.15$} eV as seen on panels (C) and (E) of Fig.~\ref{fig:cu2_pec_6panels}, but they actually worsen the Kramers-restricted GHF-CCSD results by $\sim$0.1 eV. Thus, for Cu$_2$, relativity changes the PEC by inducing a stronger Cu--Cu bond as shown by the increase in $D_\mathrm{e}$, regardless of the reference behavior.

The triples correction significantly changes the overall picture. On the Kramers-unrestricted side, we see an interesting pattern where all of the triples corrections examined in this work have similar impacts on the ground-state PECs for Cu$_2$; all triples corrections deepen the respective reference CCSD potential wells by about 0.1 eV. In contrast, the Kramers-restricted CCSD(T) and CR-CC(2,3) PECs behave differently. The CCSD(T) PECs show the well-known unphysical bumps that accompany the overcorrelation of the dissociation asymptote, regardless of the relativistic treatment used. On the other hand, all variants of CR-CC(2,3) shown in Fig.~\ref{fig:cu2_pec_6panels}(B), (D), and (F) are numerically stable and exhibit no artificial bumps along the bond dissociation coordinate. The fact that the Kramers-restricted GHF-CR-CC(2,3) PECs in Fig.~\ref{fig:cu2_pec_6panels}(B) accurately predicts the experimental $D_\mathrm{e}$ value is a coincidence; once the missing relativistic effects are added [panels (D) and (F)], the CR-CC(2,3) curves overestimate the experimentally determined $D_\mathrm{e}$ by about {\color{black} 0.16--0.20} eV. This surprising accuracy of Kramers-restricted GHF-CR-CC(2,3) suggests that errors from other effects, such as basis set size, may be responsible for a fortuitous cancellation of error. We will return to this issue in Subsection \ref{SUBSECTION:BASIS} below.

At this point, it is worth discussing how the Kramers-unrestricted and restricted PECs can help us understand the convergence of truncated CC methods toward the exact full CI limit. Because full CI is the exact solution within a given basis set, the Kramers-restricted or unrestricted HF reference must give rise to the same full CI solution. In Fig.~\ref{fig:cu2_pec_6panels}, we see that by increasing the level of electron correlation treatment, the Kramers-unrestricted and restricted potential well becomes deeper and shallower, respectively. Thus, we can use the difference between the Kramers-unrestricted and restricted energetics as an uncertainty estimate with respect to the full CI energetics. For example, in all panels, the Kramers-unrestricted GHF-, SF-1eX2C-, and 1eX2C-CCSD dissociation energies differ from their Kramers-restricted counterparts by about 0.5 eV, which indicates that CCSD is still far from converged in terms of electronic correlation effects. In contrast, the difference between the Kramers-unrestricted and restricted CR-CC(2,3) dissociation energies are on the order of 0.1--0.2 eV, which represents a massive 60\%--80\% reduction in uncertainty relative to their CCSD counterparts and thus much better convergence toward the full CI limit. The CCSD(T) approach is unfortunately not amenable to such an analysis due to the artificial bump in the 4--5 \AA~region for the Kramers-restricted Cu$_2$ PECs.

Let us now take a closer look at the other spectroscopic parameters of Cu$_2$ reported in Table \ref{tab:cu2_vdzp}. Focusing on the equilibrium bond distances, we see that CCSD(T) and all CR-CC(2,3) variants shorten the CCSD $R_\mathrm{e}$ estimates by about {\color{black} 0.01} \AA, regardless of the level of mean-field relativistic treatment. The inclusion of spin-free and spin-orbital relativistic effects is more important, shortening the equilibrium bond length by about 0.4 \AA~when compared to the non-relativistic data. This pattern also follows the expected relativistic contraction when comparing non-relativistic and relativistic calculations. Note, however, that the SF-1eX2C and 1eX2C calculations produced practically identical bond lengths, corroborating our qualitative observation of the PECs reported in Fig.~\ref{fig:cu2_pec_6panels}, which, as in the energetics analysis, indicates that spin-orbit coupling effects are minimal for Cu$_2$. We also see a similar pattern in the harmonic frequencies; the spin-free and spin-orbit-coupled 1eX2C calculations reduces the magnitude of errors relative to experiment obtained in the non-relativistic CC data, from {\color{black} 15--23} cm$^{-1}$ to about {\color{black} 1--6} cm$^{-1}$ on average. In analogy to the equilibrium bond distance, the triples correlation effects are much less pronounced than the relativistic effects, changing the harmonic frequency by only {\color{black} $\sim$6} cm$^{-1}$. Interestingly, the C and D variants of 1eX2C-CR-CC(2,3) worsen the $\omega_\mathrm{e}$ estimate compared to 1eX2C-CCSD and the A and B versions by about {\color{black} 9} cm$^{-1}$ ({\color{black} $\sim$41} microhartree). {\color{black} We observe similar pattern in the $\omega_\mathrm{e}\chi_\mathrm{e}$ values reported in Table S1 in the Supporting Information, where the C and D variants of 1e-X2C-CR-CC(2,3) produce errors that are an order of magnitude larger than those computed using the A and B variants.} This behavior could be related to the orbital invariance issues with CR-CC(2,3)$_\mathrm{C}$ and CR-CC(2,3)$_\mathrm{D}$. 

\subsection{Ag$_2$}
\label{SUBSECTION:Ag2}

\begin{figure*}[!htbp]
    \centering
    \includegraphics[width=\textwidth]{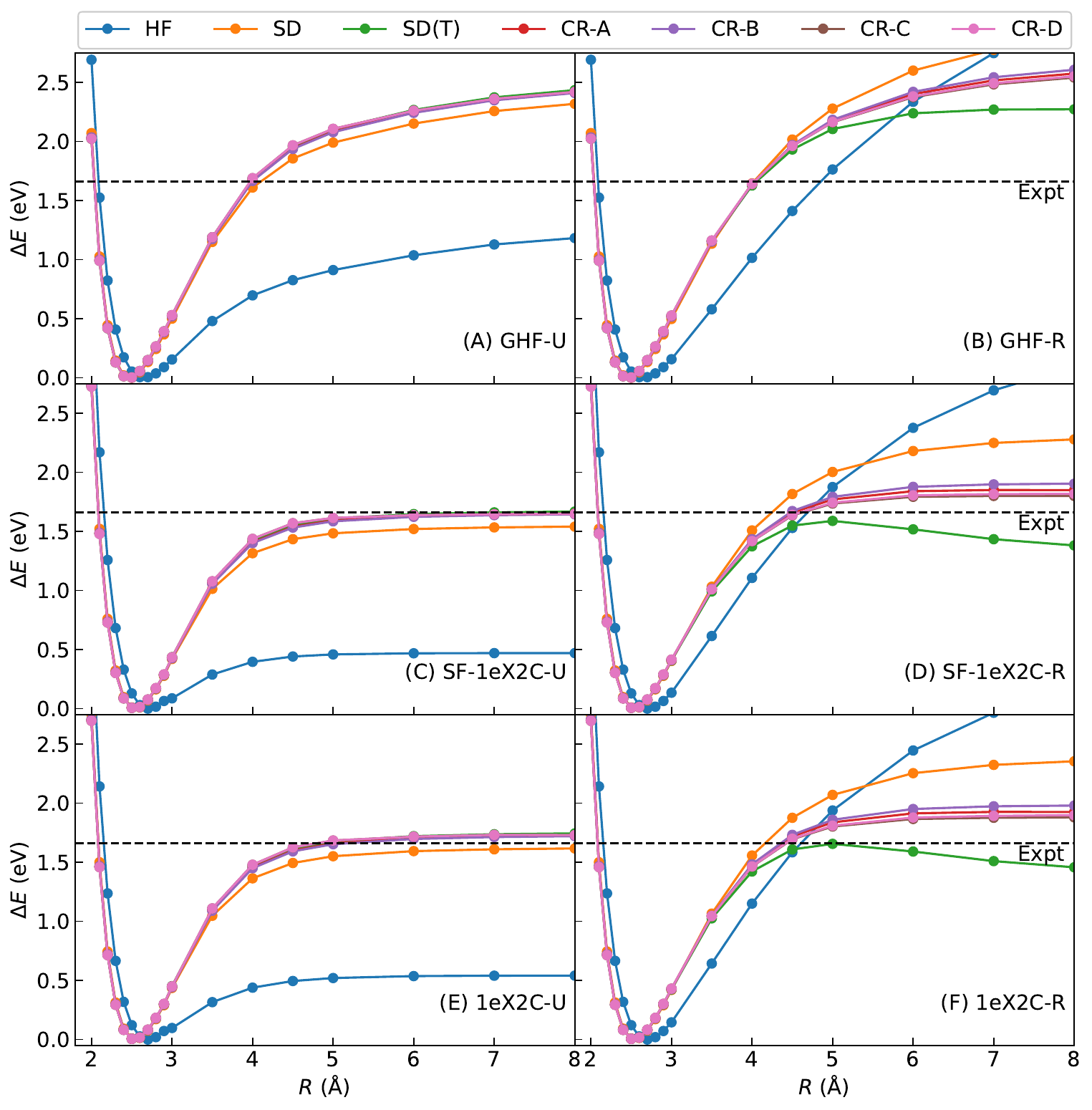}
    \caption{
        \label{fig:ag2_pec_6panels}
        The PECs of Ag$_2$ obtained in this work using the ANO-RCC-VDZP basis set. The U and R labels indicate Kramers-unrestricted and restricted reference curves, respectively. Each PEC is shifted relative to the energy at $R_\mathrm{e}$ (cf.~Table \ref{tab:ag2_vdzp}).}
\end{figure*}

\begin{sidewaystable}
\begin{minipage}{\textwidth}
    \caption{
        \label{tab:ag2_vdzp}
        Spectroscopic constants for Ag$_2$ obtained using various CC methodologies with the ANO-RCC-VDZP basis set and different levels of mean-field relativistic treatment.
        {\color{black} All values are reported as deviations from experimentally-obtained values.}}
    \centering
    {\color{black}
    \begin{tabular*}{\linewidth}{@{\extracolsep{\fill}} cccccccccc}
        \hline\hline
        \multirow{2}{*}{Method} & \multicolumn{3}{c}{$D_\mathrm{e}$ (eV)\footnotemark[1]} & \multicolumn{3}{c}{$R_\mathrm{e}$ (\AA)} & \multicolumn{3}{c}{$\omega_\mathrm{e}$ (cm$^{-1}$)} \\
        \cline{2-4} \cline{5-7} \cline{8-10}
        & GHF & SF-1eX2C & 1eX2C & GHF & SF-1eX2C & 1eX2C & GHF & SF-1eX2C & 1eX2C \\
        \hline
        CCSD                    & 0.70 (1.21) & $-0.12$   (0.62)  & $-0.04$   (0.69)  & $-0.058$ & 0.016 & 0.011 & 1.7 & $-2.6$ & $-0.6$ \\
        CCSD(T)                 & 0.82 (0.61) &   0.01  ($-0.28$) &   0.09  ($-0.20$) & $-0.069$ & 0.008 & 0.003 & 5.9 & $-1.4$ &   0.6  \\
        CR-CC(2,3)$_\mathrm{A}$ & 0.80 (0.92) & $-0.01$   (0.19)  &   0.07    (0.26)  & $-0.067$ & 0.009 & 0.004 & 5.2 & $-1.4$ &   0.6  \\
        CR-CC(2,3)$_\mathrm{B}$ & 0.80 (0.95) & $-0.01$   (0.24)  &   0.06    (0.32)  & $-0.066$ & 0.009 & 0.004 & 4.8 & $-1.3$ &   0.8  \\
        CR-CC(2,3)$_\mathrm{C}$ & 0.80 (0.88) & $-0.01$   (0.14)  &   0.07    (0.22)  & $-0.069$ & 0.007 & 0.002 & 6.8 &   0.1  &   1.6  \\
        CR-CC(2,3)$_\mathrm{D}$ & 0.80 (0.89) & $-0.01$   (0.16)  &   0.07    (0.24)  & $-0.069$ & 0.007 & 0.002 & 6.8 &   0.1  &   1.6  \\
        \hline
        Experiment\footnotemark[2] & \multicolumn{3}{c}{1.66} & \multicolumn{3}{c}{2.5303} & \multicolumn{3}{c}{192.4} \\
        \hline\hline
    \end{tabular*}
    }
    \footnotetext[1]{The numbers outside and inside parentheses refer to Kramers-unrestricted and restricted dissociation energies, respectively. The latter estimates are computed as $D_\mathrm{e} = E(R = 8.00\;\mbox{\AA})-E(R = R_\mathrm{e})$.}
    \footnotetext[2]{$D_\mathrm{e}$ computed using the $D_0$ value compiled in Ref.~\citenum{Morse86_1049} and $\omega_\mathrm{e}$ of Ref.~\citenum{Lindkvist55_385}. $R_\mathrm{e}$ from Ref.~\citenum{Demtroder93_2699}.}
\end{minipage}
\end{sidewaystable}

We proceed to the next dimer in the series, Ag$_2$, focusing on the $^{107}$Ag$_2$ isotopologue. As above, we interested in the ground $^1\Sigma_{g}^{+}$ state, which, at the dissociation limit, separates into two Ag atoms with $^2$S ([Kr] $4d^{10}5s^{1}$) electronic configurations. The ground state of this dimer, like that of Cu$_2$, has been the subject of extensive experimental and theoretical investigations (see, {\em e.g.}, Refs.~\citenum{Lindkvist55_385,Morse86_1049,Vujisic91_516,Langridge-Smith91_415,Demtroder93_2699,Balasubramanian93_7092,Fantucci99_3876,Puzzarini05_283,Liu05_63,Antic-Jovanovic08_111,Antic-Jovanovic11_786,Wang17_88,Candido21_9832}). Note that the majority of the available experimental investigation correspond to the mixed $^{107,109}$Ag$_2$ dimer, which is about twice as abundant as the $^{107}$Ag$_2$ one. Nevertheless, due to the very small ({\em i.e.}, $<1$\%) difference in the reduced masses of the two isotopologues, the impact on the numerical data of interest is negligible. The experimentally derived spectroscopic constants for Ag$_2$ are $D_\text{e}$ of about 1.66 eV ($D_0=1.65$ eV plus a zero-point energy of about 96 cm$^{-1}$ or 0.01 eV),\cite{Lindkvist55_385,Morse86_1049} $R_\mathrm{e}=2.5303$ \AA,\cite{Demtroder93_2699} and $\omega_\mathrm{e}=192.4$ cm$^{-1}$.\cite{Lindkvist55_385}
Given that the nuclear charge and mass of the $^{107}$Ag isotope are 62\% and 70\% {\color{black} higher, respectively,} than those of $^{63}$Cu, we can anticipate that relativistic effects will play a more important role in the overall energetics of of Ag$_2$. Indeed, our data, which are summarized in Fig.~\ref{fig:ag2_pec_6panels} and Table \ref{tab:ag2_vdzp}, are consistent with this expectation.

Figure~\ref{fig:ag2_pec_6panels} depicts ground-state PECs of Ag$_2$ computed using the ANO-RCC-VDZP basis set with varying levels of relativistic and electron correlation effects, while following either Kramers-unrestricted or restricted reference curves. We note the following differences between the PECs generated for Ag$_2$ and those for Cu$_2$ that we discussed above. First, consider the exaggerated attractive interaction in the GHF-based curves for Ag$_2$ shown in Fig.~\ref{fig:ag2_pec_6panels}(A) and (B). While this behavior is expected for the Kramers-restricted PEC, it is surprising to see that remnants of long-range interactions remain in the Kramers-unrestricted case, even after electronic correlation effects are included. Despite this peculiar behavior, we note that each of the Kramers-unrestricted PECs are still properly separable, {\em i.e.}, the energy at large Ag--Ag separations (100 \AA) is twice the energy of an Ag atom. The unexpected overbinding of GHF-based CCSD/CCSD(T)/CR-CC(2,3) PECs may be due to use of the ANO-RCC-VDZP basis set for non-relativistic all-electron correlated calculations. This basis set was designed for modeling valence and semicore correlations with spin-free relativistic treatments. Second, it is noteworthy that the restricted GHF-CCSD(T) curve [Fig.~\ref{fig:ag2_pec_6panels}(B)] does not show the expected unphysical hump, at least up to $R=8.00$ \AA. The lack of this feature could be attributable to the severe overbinding observed in the non-relativistic PECs.

The inclusion of relativistic effects leads to a dramatic improvement in the description of the ground-state PEC of Ag$_2$. Spin-free relativistic effects [SF-1eX2C, Fig.~\ref{fig:ag2_pec_6panels}(C)] eliminate the artificial long-range interaction seen in the Kramers-unrestricted GHF PECs in Fig.~\ref{fig:ag2_pec_6panels}(A). The quality of the Kramers-unrestricted SF-1eX2C-based PECs of Ag$_2$ in Fig.~\ref{fig:ag2_pec_6panels}(C) is comparable to that of their Cu$_2$ counterparts shown in Fig.~\ref{fig:cu2_pec_6panels}(C); the $D_\mathrm{e}$ derived from Kramers-unrestricted SF-1eX2C-CCSD differs from the experimental value by only $\sim$0.1 eV, while Kramers-unrestricted SF-1eX2C-CCSD(T) and CR-CC(2,3) provide consistent results that reduce this error by an order of magnitude. This situation contrasts with that for the Kramers-restricted case, the data for which are reported in Fig.~\ref{fig:ag2_pec_6panels}(D). Here, we find appreciable differences between the PECs obtained using the four SF-1eX2C-CR-CC(2,3) variants, especially in the $R=5.00$--8.00 \AA~region. The B variant leads to the largest deviation from the experimental $D_\mathrm{e}$ data, followed by variant A, while CR-CC(2,3)C and D produce slightly better $D_\mathrm{e}$ estimates (cf.~the SF-1eX2C-based dissociation energy values in Table \ref{tab:ag2_vdzp}). These observations are consistent with the well-known behavior of the different denominators in the CR-CC framework relying on restricted and restricted open-shell HF references (see, {\em e.g.}, Ref.~\citenum{Piecuch07_11359}).

Due to the larger size of Ag$_2$ compared to Cu$_2$, spin-orbit coupling effects are also more pronounced in this case [Fig.~\ref{fig:ag2_pec_6panels}(E) and (F)]. Indeed, the 1eX2C PECs for Ag$_2$ show a noticeable difference from their SF-1eX2C analogs in panels (C) and (D), unlike the practically identical SF-1eX2C and 1eX2C PECs obtained for Cu$_2$ (Fig.~\ref{fig:cu2_pec_6panels}). In particular, for each of the methods shown in Table \ref{tab:ag2_vdzp}, the 1eX2C-based $D_\mathrm{e}$ estimates are about 0.07--0.08 eV higher than their SF-1eX2C counterparts in both the restricted and unrestricted cases. Even though spin-orbit coupling effects are minimal for the $^1\Sigma_{g}^{+}$ state and the dissociated $^2\mathrm{S} + {}^2\mathrm{S}$ configurations, they still impact the HF orbital energies and spatial splittings,\cite{Weigend06_4862,Weigend17_3696} as well as the subsequent CC energetics. Despite the overall change in the PECs upon the inclusion of spin-orbit coupling, it is encouraging to see that the triples correction afforded by CCSD(T) and all variants of CR-CC(2,3) behave consistently as in the SF-1eX2C case; they deepen the underlying CCSD potential well by about 0.1 eV {\color{black} in the Kramers-unrestricted case. In the restricted PECs, the triples correction lowers the dissociation asymptote by about 0.4--0.5 eV, an effect that is two times larger than in Cu$_2$, which emphasizes the need for connected triples to properly describe bond-breaking in this dimer.}

As in the case of Cu$_2$, we can use the difference between Kramers-unrestricted and restricted dissociation asymptotes of Ag$_2$ as a proxy for the completeness of the correlation treatment in CCSD and CR-CC(2,3). Here, the CCSD asymptotes differ by more than 0.7 eV in both SF-1eX2C and 1eX2C cases, which is almost 50\% of the experimentally derived value of $D_\mathrm{e}$ itself. In contrast, the non-iterative triples correction significantly reduce this error bar. In the case of \mbox{(SF-)}1eX2C-CR-CC(2,3) variants A, C, and D, the difference between Kramers-unrestricted and restricted $D_\mathrm{e}$ estimate is only 0.15--{\color{black}0.20} eV, or about 9\%--{\color{black}12}\% of the experimentally derived dissociation energy. This difference is slightly larger in the case of the CR-CC(2,3)$_\mathrm{B}$ approach ({\color{black}$\sim$0.25} eV, see Table \ref{tab:ag2_vdzp}), which is still a significant improvement over CCSD. As in the case of Cu$_2$, we cannot apply a similar analysis for SF-1eX2C- and 1eX2C-CCSD(T) due to the artificial hump around 5 \AA. 

We now consider the remaining spectroscopic properties tabulated in Table \ref{tab:ag2_vdzp}. For the equilibrium distance, it is noteworthy that the $R_\mathrm{e}$ estimate for Ag$_2$ does not show the expected shortening upon the inclusion of relativistic effects. GHF-based CC predicts a bond length that is too short compared to experiment by about {\color{black} 0.06--0.07} \AA. Spin-free relativistic effects (SF-1eX2C) increase the bond length, resulting in $R_\mathrm{e}$ values that are too large by about {\color{black} 0.01} \AA. Spin-orbit coupling effects (1eX2C) in turn lead to modest reductions in the equilibrium bond lengths, by roughly {\color{black} 0.005--0.008} \AA. In terms of correlation effects, triples corrections have a smaller impact on  $R_\mathrm{e}$ for Ag$_2$ than for Cu$_2$. Similar patterns can be identified for the  $\omega_\mathrm{e}$ estimates in Table \ref{tab:ag2_vdzp}. The harmonic frequency characterizing the GHF-based PECs are 
{\color{black} already less than 10 cm$^{-1}$ higher}
than the experimentally obtained value, indicating that the PECs are slightly too 
{\color{black} curved} near the equilibrium region, whereas the SF-1eX2C and 1eX2C PECs 
{\color{black} obtained an excellent (1--2 cm$^{-1}$) agreement with experiment. The $\omega_\mathrm{e}\chi_\mathrm{e}$ values, reported in Table S2 in the Supporting Information, also show improvement in accuracy upon the inclusion of relativistic effects.}
In these cases, the triples correction changes the $\omega_\mathrm{e}$ value by only {\color{black}1} cm$^\mathrm{-1}$ {\color{black} or less}. The fact that non-relativistic calculations yield {\color{black} comparable} estimates of the spectroscopic constants, $R_e$ and $\omega_e$, {to those that} can be obtained from calculations that include more complete description of the physics suggests a cancellation of errors that is likely due to basis set size effects. This effect is not negligible for Ag$_2$, where the ANO-RCC-VDZP basis set has fewer unoccupied (78) than occupied (94) spin orbitals, indicating the need of larger basis sets for post-HF calculations, especially when all electrons are correlated.

\subsection{Au$_2$}
\label{SUBSECTION:Au2}

\begin{figure*}[!htbp]
    \centering
    \includegraphics[width=\textwidth]{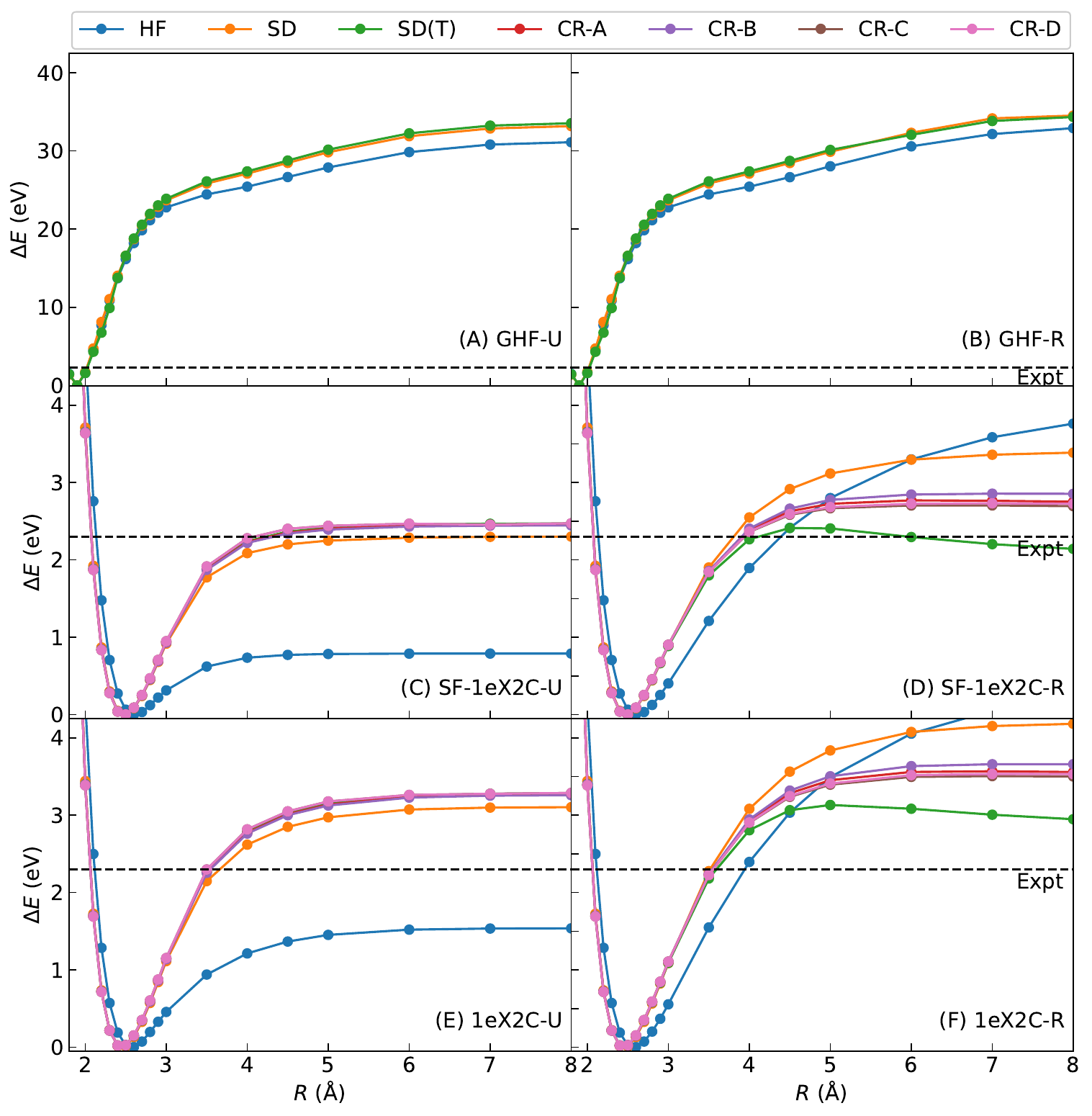}
    \caption{
        \label{fig:au2_pec_6panels}
        The PECs of Au$_2$ obtained in this work using the ANO-RCC-VDZP basis set. The U and R labels indicate Kramers-unrestricted and restricted reference curves, respectively. Each PEC is shifted relative to the energy at $R_\mathrm{e}$ (cf.~Table \ref{tab:au2_vdzp}).}
\end{figure*}

\begin{sidewaystable}
\begin{minipage}{\textwidth}
    \caption{
        \label{tab:au2_vdzp}
        Spectroscopic constants for Au$_2$ obtained using various CC methodologies with the ANO-RCC-VDZP basis set and different levels of mean-field relativistic treatment.
        {\color{black} All values are reported as deviations from experimentally-obtained values.}}
    \centering
    {\color{black}
    \begin{tabular*}{\linewidth}{@{\extracolsep{\fill}} cccccccccc}
        \hline\hline
        \multirow{2}{*}{Method} & \multicolumn{3}{c}{$D_\mathrm{e}$ (eV)\footnotemark[1]} & \multicolumn{3}{c}{$R_\mathrm{e}$ (\AA)} & \multicolumn{3}{c}{$\omega_\mathrm{e}$ (cm$^{-1}$)} \\
        \cline{2-4} \cline{5-7} \cline{8-10}
        & GHF & SF-1eX2C & 1eX2C & GHF & SF-1eX2C & 1eX2C & GHF & SF-1eX2C & 1eX2C \\
        \hline
        CCSD                    & 30.99 (32.26)       & 0.00   (1.09)  & 0.81 (1.88) & $-0.586$            & 0.007 & $-0.021$ & 779.2               & 5.7 & 16.4 \\
        CCSD(T)                 & 31.35 (32.08)       & 0.17 ($-0.16$) & 0.98 (0.64) & $-0.589$            & 0.003 & $-0.025$ & 766.6               & 5.9 & 15.2 \\
        CR-CC(2,3)$_\mathrm{A}$ & ---\footnotemark[2] & 0.16   (0.45)  & 0.97 (1.25) & ---\footnotemark[2] & 0.002 & $-0.026$ & ---\footnotemark[2] & 6.6 & 15.9 \\
        CR-CC(2,3)$_\mathrm{B}$ & ---\footnotemark[2] & 0.15   (0.56)  & 0.95 (1.35) & ---\footnotemark[2] & 0.002 & $-0.026$ & ---\footnotemark[2] & 6.7 & 16.0 \\
        CR-CC(2,3)$_\mathrm{C}$ & ---\footnotemark[2] & 0.15   (0.40)  & 0.97 (1.19) & ---\footnotemark[2] & 0.001 & $-0.027$ & ---\footnotemark[2] & 5.5 & 17.6 \\
        CR-CC(2,3)$_\mathrm{D}$ & ---\footnotemark[2] & 0.15   (0.43)  & 0.97 (1.23) & ---\footnotemark[2] & 0.001 & $-0.027$ & ---\footnotemark[2] & 5.5 & 17.6 \\
        \hline
        Experiment\footnotemark[3] & \multicolumn{3}{c}{2.30} & \multicolumn{3}{c}{2.4719} & \multicolumn{3}{c}{190.9} \\
        \hline\hline
    \end{tabular*}
    }
    \footnotetext[1]{The numbers outside and inside parentheses refer to Kramers-unrestricted and restricted dissociation energies, respectively. The latter estimates are computed as $D_\mathrm{e} = E(R = 8.00\;\mbox{\AA})-E(R = R_\mathrm{e})$.}
    \footnotetext[2]{Calculations not performed because the reference curve has a significantly wrong shape compared to the expected result.}
    \footnotetext[3]{Refs.~\citenum{Barrow67_39,Morse86_1049,Hackett90_310,Morse94_248}.}
\end{minipage}
\end{sidewaystable}

The last system in our investigation is the Au$_2$ dimer, for which we computed the ground $^1\Sigma_{g}^{+}$ state dissociating into two $^{197}$Au atoms with $^2$S ([Xe] $5d^{10} 4f^{14} 5s^{1}$) configurations. Gold dimer has attracted a great deal of attention from the quantum chemistry community because the relativistic effects are much more pronounced in Au$_2$ than in the lighter dimers discussed above. For example, the experimentally determined bond length for Au$_2$, 2.4719 \AA,\cite{Barrow67_39,Morse86_1049} is slightly shorter than the corresponding value for Ag$_2$
despite the increased size of Au atom compared to Ag atom. Au$_2$ has long served as a benchmark system for relativistic quantum chemistry calculations, and its ground-state PEC has been the subject of thorough experimental and theoretical examinations,\cite{Barrow67_39,Morse86_1049,Hackett90_310,Pitzer79_293,McLean79_288,Boyd89_1762,Lineberger90_6987,Balasubramanian90_280,Morse94_248,Schwerdtfeger99_9457,Schwerdtfeger00_9356,Kaldor00_1809,Puzzarini05_283,Liu05_63,Wang17_88,Candido21_9832} 
with the experiments providing consistent and reliable estimates of $D_\mathrm{e}$ (2.30 eV) and $\omega_\mathrm{e}$ (190.9 cm$^{-1}$),\cite{Barrow67_39,Morse86_1049,Hackett90_310,Morse94_248} along with the aforementioned $R_\mathrm{e}$ value. However, 
to the best of our knowledge, prior correlated calculations of the gold dimer have all relied on the frozen-core approximation or effective core potentials.\cite{Schwerdtfeger99_9457,Schwerdtfeger00_9356,Kaldor00_1809} As such, our calculations, the results of which are provided in Fig.~\ref{fig:au2_pec_6panels} and Table \ref{tab:au2_vdzp}, represent the first all-electron correlated relativistic treatment of this system. 

The data in Fig.~\ref{fig:au2_pec_6panels} make it quite clear that the non-relativistic results shown in panels (A) and (B) are wrong even at the qualitative level, overestimating the experimentally determined dissociation energy by a factor of about {\color{black} 15} (cf.~Table \ref{tab:au2_vdzp}), even with the inclusion of triples effects from CCSD(T). This observation is notable, given that correlated, non-relativistic estimates of $D_\mathrm{e}$ from the literature\cite{Schwerdtfeger99_9457,Schwerdtfeger00_9356} are in much better agreement with experiment. There are several possible sources of the enormous error we observe, relative to experimental and theoretical estimates for $D_\mathrm{e}$ found in the literature. First, Refs.~\citenum{Schwerdtfeger99_9457,Schwerdtfeger00_9356} do not correlate all electrons, whereas we do. Thus, we expect an explicit treatment of at least spin-free relativistic effects to be important here; we cover this case in the next paragraph. Second, we correlate all electrons, but the basis set employed (ANO-RCC-VDZP) was not optimized for all-electron calculations;\cite{Widmark05_6575} a brief basis set study in subsection \ref{SUBSECTION:BASIS} confirms that the basis set choice is indeed problematic.
Before moving on to discuss relativistic treatments of this system, we note that we did not perform GHF-based CR-CC(2,3) calculations because, based on the relative behavior of CCSD(T) and CR-CC(2,3) for Cu$_2$ and Ag$_2$, we do not expect CR-CC(2,3) to offer much of an improvement over CCSD(T) for this case.

As was observed for Ag$_2$, spin-free relativistic effects (SF-1eX2C) lead to a dramatic improvement in both the Kramers-unrestricted and restricted PECs as shown in Fig.~\ref{fig:au2_pec_6panels}(C) and (D). Indeed, at least using the ANO-RCC-VDZP basis set, the Kramers-unrestricted SF-1eX2C-CCSD PEC dissociation energy is quite close to the experimentally determined value, with the CCSD(T) and CR-CC(2,3) triples correction deepening the potential well by 0.15--0.17 eV or 6\%--7\% compared to the CCSD result. As shown on panel (D), the Kramers-restricted SF-1eX2C-based PECs are too deep compared to the experimental dissociation energy, producing errors in the order of more than 1 eV, in the case of CCSD, or about 0.4{\color{black}--0.5} eV, in the case of the CR-CC(2,3) variants. It is also noteworthy that in the case of Au$_2$, the spread between the lowest and highest CR-CC(2,3) dissociation energy estimates, provided by variants C and B, respectively, is {\color{black} 0.16} eV, which is about 30\% and 300\% higher than the analogous values reported for Ag$_2$ and Cu$_2$, respectively.

Based on the results obtained for the copper and silver dimer, and relying again on the fact that spin-orbit coupling effects should be minimal for the $^1\Sigma_{g}^{+}$ state of the gold dimer, one may have anticipated that the 1eX2C-based results would not differ significantly from the SF-1eX2C-based ones. However, this is clearly not the case for Au$_2$, within the ANO-RCC-VDZP basis. Indeed, Fig.~\ref{fig:au2_pec_6panels}(E) and (F) show {\color{black} that 
the} 1eX2C-based PECs are deeper by about 0.8 eV as compared to their SF-1eX2C counterparts shown in panels (C) and (D), regardless of the unrestricted or restricted nature of the reference curve. This substantial change in the PEC suggests that a more complete treatment of relativity is more important than high-order electron correlation effects for this system,
since the changes in the 1eX2C PECs due to triples corrections on top of CCSD are practically identical to the ones observed in the SF-1eX2C case.
More importantly, this large change again points to a potential deficiency in the basis set we have used. 

We now analyze the convergence of the different CC methods, using the difference between Kramers-unrestricted and restricted estimates of $D_\mathrm{e}$ as a proxy for convergence toward the exact limit. Regardless of the relativistic treatment used, CCSD produces an error window of about 1.1 eV, which, similar to the Ag$_2$ case, is almost half of the experimentally derived value of $D_\mathrm{e}$ itself. The CR-CC(2,3) triples corrections vastly reduce this uncertainty to only {\color{black} 0.2--0.4} eV. It is encouraging to see that, while the 1eX2C PECs show obvious problems when compared to their SF-1eX2C counterparts, the convergence behavior of CCSD, CCSD(T), and especially CR-CC(2,3) remains similar to that observed for the copper and silver dimers. Given that the correlated 1eX2C-based estimates for $D_\mathrm{e}$ lie above 3 eV, this convergence analysis suggests that remaining higher-order correlation effects (from quadruple excitaitons, for example), will not significantly improve the situation, and, thus, the issue is likely a basis set effect.

Before moving on to discuss the role of the basis set in this system, we analyze the remaining spectroscopic parameters for Au$_2$/ANO-RCC-VDZP that are reported in Table \ref{tab:au2_vdzp}. In terms of the equilibrium bond distance, the SF-1eX2C-based CCSD, CCSD(T), and CR-CC(2,3) $R_\mathrm{e}$ estimates are all only {\color{black} 0.001--0.007 \AA~}longer than the experimentally derived value of Refs.~\citenum{Barrow67_39,Morse86_1049}, whereas their 1eX2C-based counterparts are consistently about 0.02 \AA~too short. The differences among methods are more apparent in the harmonic frequency estimates, which is an indicator of the quality of the curvature of PEC near the equilibrium geometry. The SF-1eX2C-CCSD, CCSD(T), and CR-CC(2,3) PECs are characterized by $\omega_\mathrm{e}$ values that are about {\color{black} 5--7} cm$^{-1}$ {\color{black} higher} than the experimentally derived 190.9 cm$^{-1}$ estimate. In contrast, 
{\color{black} correlated calculations}
within the 1eX2C framework worsen the estimates of $\omega_\mathrm{e}$, giving values that are more than {\color{black} 15} cm$^{-1}$ too high. {\color{black} Interestingly, the SF-1eX2C- and 1eX2C-based $\omega_\mathrm{e}\chi_\mathrm{e}$ estimates for Au$_2$ (Table S3 of the Supporting Information) are comparable in terms of accuracy. For Au dimer, we see that triples corrections are less important than the relativistic effects for describing the spectroscopic parameters $R_\mathrm{e}$ and $\omega_\mathrm{e}$.}

\subsection{Basis set effects}
\label{SUBSECTION:BASIS}

The results of our CC calculations for Cu$_2$, Ag$_2$, and Au$_2$ carried out within the ANO-RCC-VDZP basis give rise to an interesting observation. As noted in Ref.~\citenum{Widmark05_6575}, the ANO-RCC basis set was not designed for core-electron correlations. Nevertheless, ANO-RCC-VDZP-based results for Cu$_2$ and Ag$_2$ are in good agreement with the available experimental data for these systems, even with all electrons correlated and spin-orbit coupling taken into account. The drastic reduction in the quality of the 1eX2C-based Au$_2$/ANO-RCC-VDZP PECs compared to their SF-1eX2C counterparts, though, certainly reveals issues with core correlations, particularly in the presence of spin-orbit coupling, when using this basis set. In this section, we investigate the effects of basis set size and contraction schemes on the convergence of 1eX2C-based HF, CCSD, and CCSD(T) energetics.

Let us begin by considering the basis set contraction (and truncation) scheme used in the calculations discussed in the preceding subsections. The ANO-RCC-VDZP basis set is a subset of the full ANO-RCC basis set, which was optimized using the Douglas--Kroll--Hess Hamiltonian and complete-active-space second-order perturbation theory accounting for valence and semicore electron correlations. The ANO-RCC-VDZP, VTZP, and VQZP basis sets are obtained by simple truncation of the full ANO-RCC basis set to obtain correlation-consistent-style contracted shells without re-optimizing the contraction coefficients. We compare this family of basis sets to the segmented contracted error-consistent basis sets of Refs.~\citenum{Weigend17_3696,Weigend20_5658}, which were optimized using 1eX2C-HF with spin-orbit coupling effects. The x2c-SVPall-2c, x2c-TZVPall-2c, x2c-TZVPPall-2c, and x2c-QZVPPall-2c bases have additional inner $p$- and $d$-type function, compared to their non-2c counterparts, to account for the proper $p$- and $d$-shell splittings.\cite{Weigend17_3696,Weigend20_5658} A comparison of the contracted atomic functions for the ANO-RCC and segmented contracted error-consistent basis sets is reported in Table \ref{tab:basis_shells}.

\begin{table}[!htbp]
\begin{minipage}{\columnwidth}
    \caption{
        \label{tab:basis_shells}
        The list of atomic basis function shells for Cu, Ag, and Au employed in this work.}
    \centering
    \begin{tabular*}{\columnwidth}{@{\extracolsep{\fill}}llll}
        \hline \hline
        Basis set & Cu ($n_o=29$) & Ag ($n_o=47$) & Au ($n_o=79$) \\
        \hline
        ANO-RCC-VDZP  & $5s4p2d1f$     & $6s5p3d1f$     & $7s6p4d2f$     \\
        ANO-RCC-VTZP  & $6s5p3d2f1g$   & $7s6p4d2f1g$   & $8s7p5d3f1g$   \\
        ANO-RCC-VQZP  & $7s6p4d3f2g1h$ & $8s7p5d3f2g1h$ & $9s8p6d4f2g1h$ \\
        \hline
        ANO-RCC\footnotemark[1] & \multicolumn{3}{c}{$10s9p8d6f4g2h$} \\
        \hline
        x2c-SVPall-2c   & $5s5p3d1f$    & $6s8p6d1f$     & $7s9p9d2f$      \\
        x2c-TZVPall-2c  & $6s7p4d1f$    & $8s9p7d1f$     & $11s11p10d2f$   \\
        x2c-TZVPPall-2c & $6s7p4d2f1g$  & $8s9p7d2f1g$   & $11s11p10d3f1g$ \\
        x2c-QZVPPall-2c & $11s9p6d4f2g$ & $13s12p9d4f2g$ & $16s16p12d7f2g$ \\
        \hline \hline
    \end{tabular*}
    \footnotetext[1]{The ANO-RCC basis sets for Cu, Ag, and Au have the same number of shells.}
\end{minipage}
\end{table}

\begin{figure*}[!htbp]
    \centering
    \includegraphics[width=\textwidth]{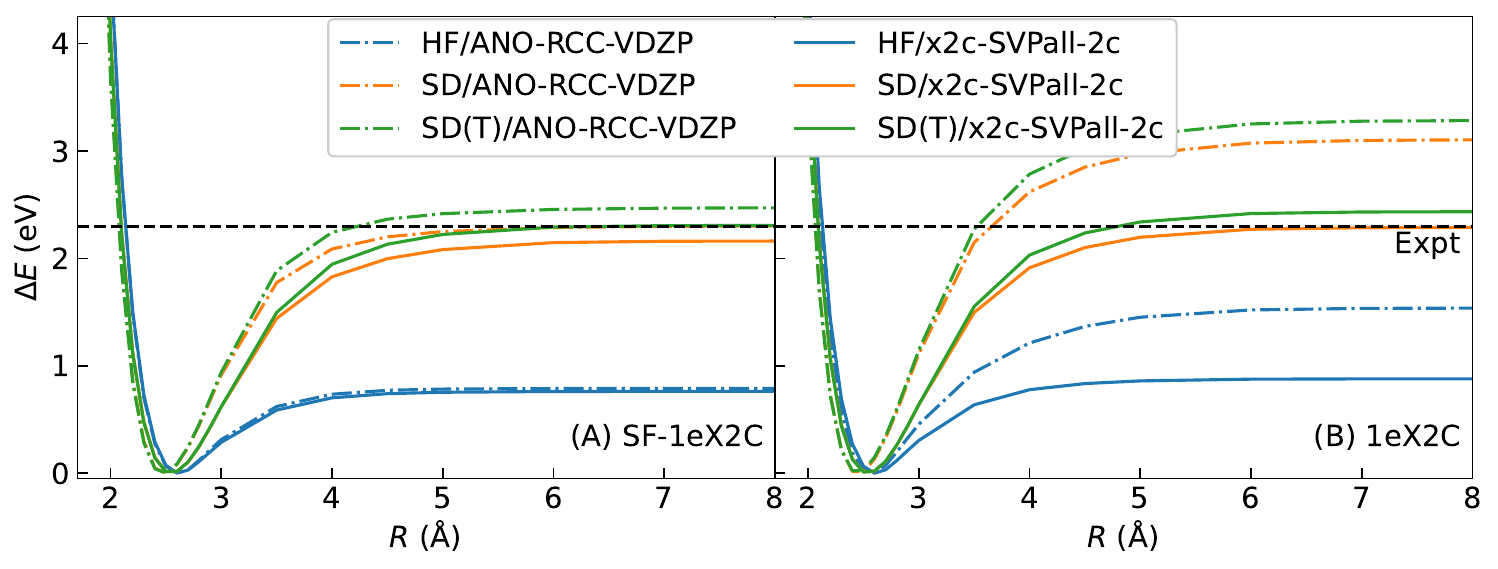}
    \caption{
        \label{fig:au2_ano_svpall}
        The Kramers-unrestricted HF, CCSD, and CCSD(T) PECs of Au$_2$ obtained in this work using the ANO-RCC-VDZP and x2c-SVPall-2c basis sets.}
\end{figure*}

Figure \ref{fig:au2_ano_svpall} depicts 1eX2C-based PECs for Au$_2$ computed using the ANO-RCC-VDZP and x2c-SVPall-2c basis sets, the latter of which is the smallest of that family of basis sets. 
As shown in Fig.~\ref{fig:au2_ano_svpall}(A), the SF-1eX2C-HF PECs are insensitive to the choice of basis set. CCSD and CCSD(T) display only slightly larger sensitivity, with the dissociation limits computed using these two basis sets differing by roughly 0.15 eV. However, as the results in panel (B) indicate, 1eX2C-HF, CCSD, and CCSD(T) energies computed in different basis sets differ dramatically once we include spin-orbit coupling effects. For the correlated calculations, we see substantial improvements in the description of the Au$_2$ PEC with the x2c-SVPall-2c  basis. The 1eX2C-CCSD and CCSD(T) $D_\mathrm{e}$ and $\omega_\mathrm{e}$ estimates are now much closer to their experimentally derived values (see Table \ref{tab:au2_ano_svpall}). Interestingly, we do not see similar improvements in equilibrium bond lengths estimated from x2c-SVPall-2c calculations; in both SF-1eX2C- and 1eX2C-based calculations, the $R_e$ values obtained using the x2c-SVPall-2c basis set are all about 0.07 \AA~longer than the experimental estimate, whereas ANO-RCC-VDZP produces values that are within 0.01--0.02 \AA~relative to experimental data.
{\color{black} Similarly, the harmonic frequency obtained using the x2c-SVPall-2c basis set are about 20 cm$^{-1}$ lower than the experimental value, showing a larger error than their ANO-RCC-VDZP counterparts, albeit without the large discrepancy between the SF-1eX2c- and 1eX2C-based data.}
Nevertheless, the improvements in {\color{black} $D_\mathrm{e}$ 
estimates} indicate that the x2c-SVPall-2c basis set is a promising alternative to ANO-RCC-VDZP.

\begin{table}[!htbp]
\begin{minipage}{\columnwidth}
    \caption{
        \label{tab:au2_ano_svpall}
        Comparison between the spectroscopic parameters of Au$_2$ obtained using the ANO-RCC-VDZP (ANO) and x2c-SVPall-2c (SVP) basis sets.
        {\color{black} All values are reported as deviations from experimentally-obtained values.}}
    \centering
    {\color{black}
    \begin{tabular*}{\columnwidth}{@{\extracolsep{\fill}}ccccccc}
        \hline \hline
        \multirow{2}{*}{Method} & \multicolumn{2}{c}{$D_\mathrm{e}$ (eV)} & \multicolumn{2}{c}{$R_\mathrm{e}$ (\AA)} & \multicolumn{2}{c}{$\omega_\mathrm{e}$ (cm$^{-1}$)} \\
        \cline{2-3} \cline{4-5} \cline{6-7}
        & ANO\footnotemark[1] & SVP & ANO\footnotemark[1] & SVP & ANO\footnotemark[1] & SVP \\
        \hline
        SF-1eX2C-CCSD    & 0.00 & $-0.14$ &   0.007  & 0.074 &  5.7 & $-20.0$ \\
        SF-1eX2C-CCSD(T) & 0.17 &   0.00  &   0.003  & 0.072 &  5.9 & $-20.6$ \\
        1eX2C-CCSD       & 0.81 & $-0.02$ & $-0.021$ & 0.069 & 16.4 & $-20.9$ \\
        1eX2C-CCSD(T)    & 0.98 &   0.13  & $-0.025$ & 0.067 & 15.2 & $-21.4$ \\
        \hline
        Experiment\footnotemark[2] & \multicolumn{2}{c}{2.30} & \multicolumn{2}{c}{2.4719} & \multicolumn{2}{c}{190.9} \\
        \hline \hline
    \end{tabular*}
    }
    \footnotetext[1]{Taken from Table \ref{tab:au2_vdzp}.}
    \footnotetext[2]{Refs.~\citenum{Barrow67_39,Morse86_1049,Hackett90_310,Morse94_248}}
\end{minipage}
\end{table}

Table \ref{tab:basis_size} summarizes the results of more comprehensive calculations that considered $D_\mathrm{e}$ values for Cu$_2$, Ag$_2$, and Au$_2$ determined using 1eX2C-HF, CCSD, and CCSD(T), within the ANO-RCC-VDZP, VTZP, VQZP, and full ANO-RCC basis sets, as well as the x2c-SVPall-2c, x2c-TZPall-2c, x2c-TZPPall-2c, and x2c-QZPPall-2c bases. In the case of Au$_2$, we also provide additional data for the ANO-RCC basis, augmented by an additional $i$-type function on each Au atom, which is prompted by a statement in Ref.~\citenum{Kaldor00_1809} claiming that the interaction in Au$_2$ can only be described properly by including at the very least $h$-type functions. For the purposes of this analysis, the $D_\mathrm{e}$ estimates are computed as the energy difference between two atoms and the lowest-energy point for each dimer,
at the CCSD/ANO-RCC-VDZP level of theory,
on the grid defined in Table \ref{tab:dist_grid}, which are $R=2.20$ \AA~for Cu$_2$ and $R=2.50$ \AA~for the Ag and Au dimers.

As shown in Table \ref{tab:basis_size}, the 1eX2C-HF/ANO-RCC-VDZP dissociation energies are practically converged for Cu$_2$ and Ag$_2$, increasing by at most 0.06 and 0.04 eV, respectively, as the basis set quality is increased from the quadruple-$\zeta$-quality truncation to the full set. Interestingly, in both of these dimers, the 1eX2C-HF dissociation energy increases from double- to {\color{black} quadruple}-$\zeta$ but then decreases once the full ANO-RCC basis set is reached, becoming more similar to the ANO-RCC-VDZP or VTZP estimates. The behavior is more unpredictable for Au$_2$, where the 1eX2C-HF dissociation energy estimate shows no clear convergence pattern. With the full ANO-RCC basis set, the $D_\mathrm{e}$ estimate is 0.85 eV, which is slightly more than half of the value obtained using ANO-RCC-VDZP. The addition of an $i$-type primitive to each of the Au atom does not significantly change the HF energy, indicating that the full ANO-RCC basis set can be considered converged in terms of 1eX2C-HF energetics for the gold dimer (as well as the lighter analogs).

The $D_\mathrm{e}$ estimates from correlated approaches show much worse convergence properties. Going from triple- to quadruple-$\zeta$-quality basis, the $D_\mathrm{e}$ is clearly not converged, and massive reductions are observed once the full set is used, again, bringing the estimates more in line with those from double- or triple-$\zeta$-level calculations. For Cu$_2$, $D_\mathrm{e}$ obtained using CCSD increase monotonically from 1.89 eV (ANO-RCC-VDZP) to 2.20 eV (ANO-RCC-VQZP) case, but a much lower value (1.77 eV) is obtained with the full ANO-RCC set. The CCSD(T) energetics behave similarly, but the correlation effects due to connected triple excitations, quantified her as the difference between CCSD(T) and CCSD energetics, more than double when going from ANO-RCC-VDZP (0.12 eV) to the full ANO-RCC (0.28 eV). It is also worth mentioning that 1e-X2C-CCSD(T)/ANO-RCC predicts a $D_\text{e}$ value in excellent agreement with experiment. The  CCSD(T)/ANO-RCC-VDZP estimate is also quite good but results from a fortuitous cancellation of error, given that the basis is far from complete. The situation is similar for the silver dimer, in which the CCSD and CCSD(T) $D_\mathrm{e}$ estimates increase as the basis set size also increases from ANO-RCC-VDZP to VQZP, and finally dropping at the use of the full ANO-RCC basis. The effects of triples correlations on the dissociation energy of Ag$_2$ are also of the similar order of magnitude to those observed in Cu$_2$, ranging from 0.13 eV in the ANO-RCC-VDZP case to 0.21 eV in the full ANO-RCC set. Unlike in the case of Cu$_2$, unconverged 1eX2C-CCSD and CCSD(T) estimates obtained using the ANO-RCC-VDZP basis are not in good agreement with the experimentally derived value. 

The basis set convergence issues for correlated approaches  persist in the gold dimer, where increasing the basis from ANO-RCC-VDZP to ANO-RCC-VTZP makes the $D_\mathrm{e}$ value worse, as compared to experiment, by 0.64 eV, and the ANO-RCC-VQZP basis set produces a result that is intermediate in quality between the ANO-RCC-VDZP and VTZP values. Only the full ANO-RCC set results in a reasonable dissociation energy value from CCSD (2.16 eV), which is an 0.93 eV or 30\% decrease from the estimate obtained using ANO-RCC-VDZP. The addition of $i$-type functions does not significantly affect the correlated estimates of $D_\mathrm{e}$, changing the CCSD value by merely 0.01 eV. The convergence of $D_\mathrm{e}$ with  respect to the addition of an $i$-type primitive is consistent with the claim in Ref.~\citenum{Kaldor00_1809} that one should to include at least $h$-type functions in the calculation; apparently higher angular momentum functions are not required, at least for obtaining good estimates of $D_\mathrm{e}$. This consistency comes with the caveat that the present calculations correlate all electrons, whereas those in Ref.~\citenum{Kaldor00_1809} correlated only valence and semicore correlation, in combination with with effective core potentials. Although we did not complete the CCSD(T)/ANO-RCC and CCSD(T)/ANO-RCC+$i$ calculations for Au$_2$, as indicated by the missing numbers in Table \ref{tab:basis_size}, the triples energy corrections for the ANO-RCC-VDZP--ANO-RCC-VQZP cases, of 0.17--0.25 eV, are still similar in magnitude to their copper and silver dimer counterpart. Thus, we can anticipate these CCSD(T) $D_\mathrm{e}$ estimates to be at most 0.2 eV higher than the experimental result. 

We now turn our attention to the convergence of the $D_\mathrm{e}$ estimates computed within the segmented contracted error-consistent basis sets of Refs.~\citenum{Weigend17_3696,Weigend20_5658} (see Table \ref{tab:basis_size}). 
First, the $D_\mathrm{e}$ values from 1eX2C-HF carried out within these basis sets converge rapidly for Cu$_2$ and Ag$_2$. For Au$_2$, we observe slightly larger variations in $D_\mathrm{e}$ as we increase the $\zeta$-level (up to 0.06 eV), but these fluctuations are minuscule compared to those observed for Au$_2$ with truncated ANO-RCC basis sets. Moreover, the quality of the 1eX2C-HF $D_\mathrm{e}$ from the smallest segmented contracted error-consistent basis set (x2c-SVPall-2c ) is comparable to that of the $D_\mathrm{e}$ from 1eX2C-HF in the full ANO-RCC set. Second, the situation is similar for correlated calculations. The $D_\mathrm{e}$ values converge reasonably well; for each dimer, $D_\mathrm{e}$ estimates do not change by more than about 0.2 eV when we go from the smallest to largest $\zeta$ levels. Third, aside from the difference in convergence properties, we do find some consistent behavior between basis set families. For example, the triples contributions to $D_\mathrm{e}$ computed using CCSD(T) are comparable regardless of the basis (on the order of 0.15--0.25 eV). Lastly, the 1eX2C-CCSD(T)/x2c-QZVPPall-2c results for the Cu$_2$ and Ag$_2$ are are in excellent agreement with experiment.  For Au$_2$, given that (i) the CCSD(T) triples correction in the x2C-TZPPall-2C basis set is $\sim$0.2 eV
and (ii) the $D_\mathrm{e}$ value computed using CCSD appears to be converged using the quadruple-$\zeta$ basis, we can expect that 1eX2C-CCSD(T)/x2c-QZVPPall-2c result should be within 0.1 eV from the experimental estimate of 2.30 eV.

\begin{sidewaystable}
\begin{minipage}{\textwidth}
    \caption{
        \label{tab:basis_size}
        The effect of basis set size and contraction scheme on the Kramers-unrestricted 1eX2C-HF-based $D_\mathrm{e}$ estimate of Cu$_2$, Ag$_2$, and Au$_2$, computed as the difference between the energy of two atoms ({\em i.e.}, the asymptote) and the lowest energy on the grid described in Table \ref{tab:dist_grid}.}
    \centering
    \begin{tabular*}{\linewidth}{@{\extracolsep{\fill}} cccccccccc}
        \hline \hline
        \multirow{2}{*}{Basis set} & \multicolumn{3}{c}{Cu$_2$, $R=2.20$ \AA} & \multicolumn{3}{c}{Ag$_2$, $R=2.50$ \AA} & \multicolumn{3}{c}{Au$_2$, $R=2.50$ \AA} \\
        \cline{2-4} \cline{5-7} \cline{8-10}
         & HF & CCSD & CCSD(T) & HF & CCSD & CCSD(T) & HF & CCSD & CCSD(T) \\
        \hline
        ANO-RCC-VDZP    & 0.42 & 1.89 & 2.01 & 0.42 & 1.61 & 1.74 & 1.51 & 3.09 & 3.26 \\
        ANO-RCC-VTZP    & 0.45 & 2.10 & 2.33 & 0.43 & 1.87 & 2.05 & 1.73 & 3.73 & 3.96 \\
        ANO-RCC-VQZP    & 0.48 & 2.20 & 2.45 & 0.46 & 2.17 & 2.37 & 1.18 & 3.49 & 3.74 \\
        ANO-RCC         & 0.44 & 1.77 & 2.05 & 0.38 & 1.85 & 2.06 & 0.85 & 2.16 & ---\footnotemark[1]  \\
        ANO-RCC+$i$     & ---\footnotemark[1]  & ---\footnotemark[1]  & ---\footnotemark[1]  & ---\footnotemark[1]  & ---\footnotemark[1]  & ---\footnotemark[1]  & 0.85 & 2.17 & ---\footnotemark[1]  \\
        \hline
        x2c-SVPall-2c   & 0.45 & 1.97 & 2.15 & 0.36 & 1.65 & 1.81 & 0.81 & 2.28 & 2.42 \\
        x2c-TZVPall-2c  & 0.42 & 1.71 & 1.92 & 0.33 & 1.43 & 1.60 & 0.76 & 1.97 & 2.17 \\
        x2c-TZVPPall-2c & 0.43 & 1.70 & 1.93 & 0.35 & 1.42 & 1.61 & 0.82 & 2.04 & 2.27 \\
        x2c-QZVPPall-2c & 0.43 & 1.81 & 2.06 & 0.34 & 1.44 & 1.65 & 0.81 & 2.07 & ---\footnotemark[1] \\
        \hline
        Experiment\footnotemark[2] & \multicolumn{3}{c}{2.02} & \multicolumn{3}{c}{1.66} & \multicolumn{3}{c}{2.30} \\
        \hline \hline
    \end{tabular*}
    \footnotetext[1]{Calculations not performed.}
    \footnotetext[2]{See footnote b in Tables \ref{tab:cu2_vdzp}--\ref{tab:au2_vdzp}.}
\end{minipage}
\end{sidewaystable}

\section{Conclusions}

\label{SEC:CONCLUSIONS}

We have implemented all-electron relativistic (1eX2C) non-iterative triples corrections to CCSD, namely, CCSD(T) and CR-CC(2,3), in the Chronus Quantum software package. These codes have been applied to evaluate estimates of spectroscopic constants in the coinage metal dimers, Cu$_2$, Ag$_2$, and Au$_2$. Using suitable basis sets, $D_\text{e}$  estimates are in excellent agreement with experiment, with triples correlation effects contributing 0.1--0.2 eV. While ANO-RCC sets were not optimized for all-electron calculations, our calculations on Cu$_2$ and Ag$_2$ dimers nonetheless give reasonable results as compared to experiment. On the other hand, calculations on Au$_2$ reveal that truncated ANO-RCC basis sets do not provide a reliable description of spin-orbit coupling effects in this system. Other basis set families optimized for all-electron calculations with spin orbit coupling ({\em i.e.}, the segmented contracted error-consistent basis sets of Refs.~\citenum{Weigend17_3696,Weigend20_5658}), appear to give more consistent and reasonable results at varying $\zeta$-levels, at least for the systems studied in this work.

\vspace{0.5cm}

\vspace{0.5cm}

\noindent {\bf ACKNOWLEDGMENTS:} \\ \\

This material is based upon work supported by the U.S.~Department of Energy (DOE), Office of Science, Office of Advanced Scientific Computing Research and Office of Basic Energy Sciences, Scientific Discovery through the Advanced Computing (SciDAC) program under Award No.~DE-SC0022263. This project used resources of the National Energy Research Scientific Computing Center, a DOE Office of Science User Facility supported by the Office of Science of the U.S.~DOE under Contract No.~DE-AC02-05CH11231 using NERSC award ERCAP-0024336, and resources of the Argonne Leadership Computing Facility, a U.S.~DOE Office of Science user facility at Argonne National Laboratory and is based on research supported by the U.S.~DOE Office of Science -- Advanced Scientific Computing Research Program, under Contract No.~DE-AC02-06CH11357. {\color{black} The development of the Chronus Quantum computational software is supported by the Office of Advanced Cyberinfrastructure, National Science Foundation (Grants No. OAC-2103717 to XL, OAC-2103705 to AED, and OAC-2103738 to EFV)}.

\bibliography{main}

\end{document}